\documentclass{emulateapj}

\usepackage{natbib} 
\usepackage{longtable} 
\usepackage{amsmath} 
\usepackage{color} 
\def\gtorder{\mathrel{\raise.3ex\hbox{$>$}\mkern-14mu
\lower0.6ex\hbox{$\sim$}}}
\def\ltorder{\mathrel{\raise.3ex\hbox{$<$}\mkern-14mu
\lower0.6ex\hbox{$\sim$}}}  
\def\bfy{{\bf y}} 
\def\bfn{{\bf n}} 
\def\bfs{{\bf s}} 
\def\bfq{{\bf q}} 
\def\bfp{{\bf p}}

\def\bfhs{{\bf\hat{s}}} 
\def\bfhq{{\bf\hat{q}}}
\def\bfds{{\bf\Delta s}} 
\def\bfdq{{\bf\Delta q}} 
\makeatletter

\newcommand{\Rmnum}[1]{\expandafter\@slowromancap\romannumeral #1@}
\makeatother

\shorttitle{Measuring Reverberation Lags}
\shortauthors{Zu et al.}

\begin{document}
\title{An Alternative Approach To Measuring
Reverberation Lags in Active Galactic Nuclei} 

\author{Ying~Zu,\altaffilmark{1} 
C.S.~Kochanek,\altaffilmark{1,2}
and 
Bradley M.~Peterson\altaffilmark{1,2}}
\altaffiltext{1}{Department of Astronomy, The Ohio State University, 140 West
18th Avenue, Columbus, OH 43210; yingzu@astronomy.ohio-state.edu.}
\altaffiltext{2}{Center for Cosmology and AstroParticle Physics, The Ohio
State University, 191 West Woodruff Avenue, Columbus, OH 43210}
\begin{abstract}
Motivated by recent progress in the statistical modeling of quasar
variability, we develop a new approach to measuring emission-line
reverberation lags to estimate the size of broad-line regions~(BLRs)
in active galactic nuclei. Assuming that all emission-line light
curves are scaled, smoothed, and displaced versions of the continuum,
this alternative approach fits the light curves directly using a
damped random walk model and aligns them to recover the time lag and
its statistical confidence limits. We introduce the mathematical
formalism of this approach and demonstrate its ability to cope with
some of the problems for traditional methods, such as irregular
sampling, correlated errors, and seasonal gaps. We redetermine the
lags for 87 emission lines in 31 quasars and reassess the BLR
size--luminosity relationship using 60 H$\beta$ lags. We confirm the
general results from the traditional cross-correlation methods, with
a few exceptions. Our method, however, also supports a broad range of
extensions. In particular, it can simultaneously fit multiple lines
and continuum light curves which improves the lag estimate for the
lines and provides estimates of the error correlations between them.
Determining these correlations is of particular importance for
interpreting emission-line velocity--delay maps. We can also include
parameters for luminosity-dependent lags or line responses.  We use
this to detect the scaling of the BLR size with continuum luminosity
in NGC~5548.
\end{abstract}
\section{Introduction} 
\label{sec:intro}
While it is widely accepted that the enormous luminosities of active
galactic nuclei~(AGNs) are attributable to accretion of matter onto
supermassive black holes~(BH), detailed studies are extremely
challenging on account of the small angular scales of the regions
involved in the accretion process. Direct probes of the
sub-microarcsecond structure of AGNs has been therefore limited to
VLBI studies of the radio-emitting regions, gravitational
microlensing studies of the accretion disk~(see review by
\citealt{wambsganss06}) and reverberation mapping of the broad-line
regions~(\citealt{BM82,peterson93}). The technique of reverberation
mapping (a.k.a.\ echo mapping) exploits the light travel time between
the central engine and the broad-line region~(BLR) to deduce the
structure of the BLR (see \citealt{peterson01} for a tutorial).  The
continuum radiation from the accretion disk photoionizes gas clouds
near the AGN to produce broad emission lines, thus encoding the
geometry and kinematics of the
clouds~(\citealt{osterbrock89,peterson97,krolik99}). The physical
{\it ansatz} for reverberation mapping is straightforward:
\begin{enumerate}
\item The continuum emission of the quasar shows (stochastic)
    variability that drives emission-line variations after a light
    travel-time delay.
\item The unobservable ionizing UV continuum that drives the emission
    lines is simply related to the observable satellite UV or optical
    continuum (i.e., the pattern and phase of variations are closely
    correlated).
\item The light-travel time is the most important time scale;
    specifically, the local emission-line response time to continuum
    changes is assumed to be instantaneous and the dynamical time
    scale of the BLR is much larger than the light-travel time across
    it.
\end{enumerate}

The relationship between the observables, continuum light curve
$s_c(t)$ and the emission-line light curve $s_l(t,V)$ where $V$ is
the line-of-sight velocity, is taken to be
\begin{equation}
\label{eq:RMdefn}
s_l(t, V) = \int d\tau\, \Psi(\tau, V) s_c(t -\tau),
\end{equation}
where $\Psi(\tau,V)$ is known as the ``transfer function'' or
``velocity--delay map.'' In reality, the relationship between the
continuum and emission-line variations can be non-linear, but the
amplitude of variation on reverberation time scales is sufficiently
small that the linear approximation seems to be justified.
Inspection of Equation (\ref{eq:RMdefn}) shows that $\Psi(\tau, V)$
is the observed response of the broad emission-line region to a
delta-function continuum outburst, mapped into the observable
quantities time delay $\tau$ and line-of-sight velocity $V$. The data
requirements for successful recovery of the transfer function are
quite demanding \citep{horne04} and consequently most efforts to date
have concentrated on measuring only the {\em total} emission-line
response to continuum variations.  The transfer equation (eq.
\ref{eq:RMdefn}) then becomes
\begin{equation}
\label{eq:TFdefn}
s_l(t) = \int d\tau\, \Psi(\tau) s_c(t -\tau),
\end{equation}
where $\Psi(t) = \int \Psi(\tau, V)\,dV$ is variously known as the
``one-dimensional transfer function'' (so that  $\Psi(t, V)$ is the
``two-dimensional transfer function'' or the ``delay map''). For the
remainder of this paper, we will refer to $\Psi(t)$ simply as the
``transfer function.'' In most investigations to date, it is the mean
response time or ``lag'' $\langle \tau \rangle$ that one tries to
measure, generally by cross-correlation of the continuum and
emission-line light curves, as we discuss further below. The
importance of measuring the emission-line lag is two-fold: first,
$\langle \tau \rangle$ yields a characteristic physical scale for
emission of a particular line, $R = c\langle \tau \rangle$, and this
can be combined with some measure of the emission-line Doppler width
$\Delta V$ to obtain an estimate of the central black hole mass.
Assuming that gravity is the dominant force on the line-emitting
gas, the virial equation for the central black hole mass is
\begin{equation}
\label{eq:virial}
M_{\rm BH} = \frac{f \Delta V^2 R}{G},
\end{equation}
where $G$ is the gravitational constant and $f$ is a dimensionless
factor of order unity that depends on the geometry, velocity field,
and inclination of the BLR. We note in passing that there is
currently an active debate about the relative importance of radiation
pressure on the BLR gas and how this affects reverberation-based mass
measurements (\citealt{marconi08,netzer09,marconi09,NM10}).  While
the possible role of radiation pressure in measurement of black hole
masses is an important issue, it has no direct bearing on the present
discussion, which is about measuring time delays. Similarly, there is
still active discussion about the mean value of the scaling factor
$\langle f \rangle$ (e.g., \citealt{onken04,
labita06,woo10,graham11}) that is beyond the scope of this
contribution. Second, reverberation studies have established a tight
empirical relationship between the BLR size and the AGN continuum
luminosity (\citealt{kaspi96, kaspi00, kaspi05, bentz06a, bentz09a})
that allows us to use the luminosity as a surrogate for the BLR size
in eq.\ (\ref{eq:virial}) and thus estimate the masses of black holes
in AGNs from individual spectra (\citealt{wandel99, vestergaard02,
mclure02, mclure04, vestergaard06, kollmeier06, shen08}). This allows
us to explore  BH properties and evolution with redshift~(e.g.,
\citealt{kollmeier06,peng06,HH06,shankar09,SE10,kelly10}), thus
providing valuable insights into the mystery of black hole growth and
its connection to galaxy evolution at high redshift, where the quasar
population is evolving dramatically.  The potential for obtaining
simple estimates for the masses of black holes in quasars provide a
means of exploring the correlations between the BH mass and global
properties of their host galaxies such as the bulge
luminosity~($M_{\rm BH}$--$L_{\rm bulge}$ relationship;
\citealt{KR95,magorrian98,bentz09a}) and bulge stellar velocity
dispersion~($M_{\rm BH}$--$\sigma_\star$ relationship;
\citealt{FM00}~Gebhardt et al.
2000a,b;~\citealt{tremaine02,ferrarese01,onken04,nelson04,gultekin09})
both locally and potentially over cosmic time.

As a practical problem in aperiodic time-series data analysis,
reverberation mapping requires high-fidelity spectroscopic monitoring
of the continuum and emission-line variations for a duration long
compared to the emission-line lag~\citep{horne04}, which is observed
to range from hours to a year or more, depending on the luminosity of
the AGN and the cosmic time dilation at its redshift.  Emission-line
lags have been measured for more than three dozen AGNs by cross
correlation of the continuum and emission-line light curves. The
particular challenge of dealing with reverberation time series is
that they are generally irregularly sampled for various reasons,
including unfavorable weather and, for higher-luminosity objects with
larger lags, annual conjunctions with Sun that cause seasonal gaps in
the observed time series. In practice, two methodologies have been
widely employed to deal with unevenly sampled data.  The first method
is to interpolate between real data points to obtain a regular
sampling grid for computation of the cross-correlation function (CCF)
as a function of time delay $\tau$ (\citealt{GS86, GP87, WP94,
peterson98, welsh99, peterson04}).  The second method, the discrete
correlation function (DCF) method \citep{EK88}, bins the data over
discrete time intervals on which the data are reasonably well-sampled
and a correlation coefficient is computed for the time delays between
each pair of continuum/emission-line time bins. A variant on this is
the $Z$-transformed DCF \citep{alexander97} which varies the width of
the time bins to better distribute the data points among the time
bins.  \cite{WP94} show that when common assumptions and
normalizations are used, the interpolation CCF method and the DCF
method give similar results. However, as the time-sampling becomes
sparser, the interpolation method significantly outperforms the DCF
method as long as interpolation of the light curves (usually linear
in practice) remains a reasonable assumption\footnote{Consider, as an
example, the UV and optical monitioring campaign on NGC 5548
undertaken with the {\em International Ultraviolet Explorer}
\citep{clavel91} and ground-based telescopes \citep{peterson91} in
1989. The UV data were sampled at approximately regular 4-day
intervals. Analysis of these data using the interpolation CCF
\citep{PW99} revealed for the first time a ``virial relationship''
between the emission-line lags and line widths (i.e., $\langle \tau
\rangle \propto \Delta V^{-2}$), thus providing an empirical
justification for using Equation (\ref{eq:virial}) to estimate the
black hole mass. Analysis of these same data with the DCF method
\citep{krolik91} obscured this result at least in part because of
``discretization noise'' introduced by the DCF: because of the
regular 4-day sampling, the smallest usable time bin for the DCF was
also four days, resulting in lag measurements that were integer
multiples of 4 days, significantly reducing the time resolution of
the lag measurements and smearing out the $\langle \tau
\rangle$--$\Delta V$ anticorrelation.}.

Presumably even more accurate lags could be
measured given more realistic modeling of the continuum behavior
between real measurements of the continuum and emission-line fluxes.
This now seems to be a real possibility given the recent work of
\cite{kelly09}, who find that quasar variability can be well
described by a damped random walk. By applying the variability model
to the light curves of known quasars and comparing them to other
variable sources, \cite{koz10a} show that quasars occupy a very
distinctive region in the model parameter space of time scale and
variability amplitude. \cite{macleod10} then apply the model to
$\sim$ 9,000 spectroscopically identified quasars in SDSS Stripe 82.
They confirm that the model can describe quasar variability well and
they explore the correlation of the variability parameters with other
properties of quasars such as wavelength, luminosity, BH mass, and
Eddington ratios in detail. Importantly for reverberation studies,
the formalism is able to statistically predict the value of light
curve at an unmeasured time based on the overall statistical
properties of the light curve. It provides a well-defined statistical
model for interpolating light curves and can do appropriate
statistical averages over the uncertainties in the model predictions.

Given a complete statistical framework for describing the continuum
variability, and the overall {\it ansatz} that emission-line
variability is a scaled and smoothed version of the continuum, we can
build an alternative approach to measuring reverberation lags,
aspects of which were previously noted by \cite{RK94}.  Among the
advantages of this approach are:
\begin{enumerate}
\item It not only interpolates between data points, but also
    self-consistently estimates and includes the uncertainties in the
    interpolation.
\item It can separate light curve means, trends, and systematic
    errors in flux calibration from variability signals and
    meansurement noise in a self-consistent way. 
\item Correlated errors can be treated naturally.
\item Lags of multiple emission lines and their covariances can be
    derived simultaneously.
\item It provides statistical confidence limits on the lag estimates
    as well as other parameters. 
\end{enumerate}
We describe the methodology of our approach in detail in
\S\ref{sec:method}. In \S\ref{sec:cont}, we present the statistical
process model for the continuum light curves. We briefly describe our
data set and apply this method to the estimate of H$\beta$ lags in
\S\ref{sec:hbeta}. We further show how the method can address the
problem of correlated errors in \S\ref{sec:corr} and how it can be
used to improve lag estimates, particularly in the presence of
seasonal gaps, by fitting multiple lines simultaneously in
\S\ref{sec:doublehat}. We also fit the $R_{\rm BLR}$--$L$
relationship using H$\beta$ lags determined by our method in
\S\ref{sec:RLrelation}.  In \S\ref{sec:breath}, we add a luminosity
dependence to the lag and solve for the lag--luminosity relationship
of NGC~5548. We summarize our main findings and discuss future
applications and expansions of our approach in \S\ref{sec:dis}. 

%
\section{Methodology} \label{sec:method}
\cite{PRH92} and \cite{RP92} developed a method to statistically
analyze irregularly sampled light curves, and \cite{RK94} applied the
variant we now consider to four seasons of optical reverberation data
on NGC~5548. Here we reintroduce this approach, which we have named
``Stochastic Process Estimation for AGN Reverberation
(SPEAR\footnote{\url{http://www.astronomy.ohio-state.edu/\string~yingzu/spear.html}}),''
with several modest changes in algorithm and a broad range of new
applications.

Except for the transfer function $\Psi(\tau)$, our notation is chosen
for comparison with \cite{RK94}.   We start with a model process
driving the continuum $s_c(t)$ that has a covariance between times
$t_i$ and $t_j$ of 
\begin{equation}
      \langle s_c(t_i) s_c(t_j)\rangle = \sigma^2 \exp(-|t_i-t_j|/\tau_{\rm d}).
      \label{eqn:cfunc}
\end{equation}
We adopt here an exponential covariance matrix for concreteness,
since we know from \cite{kelly09}, \cite{koz10a} and \cite{macleod10}
that quasar light curves are well modeled by this process.
Physically, the model corresponds to a random walk described by an
amplitude $\sigma^2=\hat{\sigma}^2\tau_{\rm d}/2$ on long time scales
and an exponential damping time scale $\tau_{\rm d}$, where
$\hat{\sigma}$ and $\tau_{\rm d}$ are used as our model parameters.
\cite{RP92} estimated the covariance matrix based on the structure
function of the continuum light curve, while here we adopt a specific
parametrized model that will be optimized as part of the analysis.

Slightly rewriting Equation (\ref{eq:TFdefn}) for convenience and to
facilitate comparison with \cite{RK94}, the light curve of a line is
\begin{equation}
      s_l(t) \equiv \int dt' \Psi(t-t') s_c(t).
\end{equation}
Since the lines and continuum are related by the transfer function,
we can also determine the covariance between the line and continuum
\begin{equation}
    \langle s_l(t_i) s_c(t_j)\rangle = \int dt' \Psi(t_i-t') \langle  s_c(t')
    s_c(t_j) \rangle,
    \label{eqn:lc}
\end{equation} 
between the line and itself
\begin{equation}
    \langle s_l(t_i) s_l(t_j)\rangle = \int dt' dt'' \Psi(t_i-t') \Psi(t_j-t'')\langle
    s_c(t') s_c(t'') \rangle,
    \label{eqn:lauto}
\end{equation}
and between two different lines
\begin{equation}
    \langle s_l(t_i) s_l'(t_j)\rangle = \int dt' dt'' \Psi(t_i-t')
    \Psi'(t_j-t'')\langle  s_c(t') s_c(t'') \rangle.
    \label{eqn:lcross}
\end{equation}
If the light curve of the line is divided into velocity bins $\delta
V$, then there is a transfer function for each bin $\Psi(t-t', V)$
and we can compute all the expected covariances between the light
curves. For convenience, let $ \bfs$ be a vector comprised of all the
light curves, both line and continuum, and $S = \langle \bfs \bfs
\rangle$ be the covariance matrix between all the elements of $\bfs$.
By definition, in Gaussian statistics the probability of the light
curve is simply
\begin{equation}
   P(\bfs) \propto \left| S \right|^{-1/2} \exp\left(  - { \bfs^T S^{-1} \bfs
   \over 2 } \right).
\end{equation}
We do not measure the actual light curve, but some realization of it,
$\bfy = \bfs + \bfn + L \bfq$, in which there is measurement error
$\bfn$, whose probability distribution is
\begin{equation}
   P(\bfn) \propto \left| N \right|^{-1/2} \exp\left(  - { \bfn^T N^{-1} \bfn
   \over 2 } \right).
\end{equation}
where $N = \langle \bfn \bfn \rangle$ is the covariance matrix of the
noise.  Note that nothing requires $N$
to be diagonal, so there is no formal difficulty to including
covariances in the noise between the line and continuum.
\begin{figure}[t]
\epsscale{1.0} \plotone{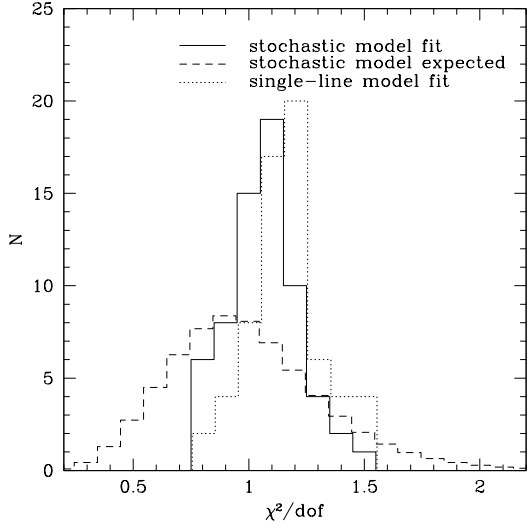}
\caption{Distribution of $\chi^2$ per degree of freedom for the
continuum fits. The solid histogram is the $\chi^2/dof$ distribution
of our stochastic model, while the dashed one shows the distribution
expected for models with correctly estimated Gaussian uncertainties.
The dotted histogram is the $\chi^2/dof$ distribution of the joint
model of the continuum and H$\beta$ light curves from
\S\ref{sec:hbeta}.}
\label{fig:chi2dof}
\end{figure}

In defining $\bfy$, we have also allowed for the simultaneous fitting
of a general trend defined by a response matrix $L$ and a set of
linear coefficients $\bfq$. In particular, we use this to fit and
remove separate means from the light curves. In this application to a
model with two light curves, $L$ is a $2 \times K$ matrix with
entries of $(1,0)$ for the continuum data points and $(0,1)$ for the
line data points, where $K$ is the total number of data points. The
linear parameters are a very general tool. For example, separate
linear trends would be removed with a $4 \times K$ matrix with
entries of $(1,t_i,0,0)$ for continuum epoch $t_i$ and $(0,0,1,t_j)$
for line epoch $t_j$. Two sources of data with potentially different
{\it but constant} levels of contamination from the host galaxy can
be reconciled by using different means for each line and continuum
data source, corresponding to a $4\times K$ matrix with entries of
$(1,0,0,0)$ for the first continuum source, $(0,1,0,0)$ for the
second continuum source, $(0,0,1,0)$ for the first line source and
$(0,0,0,1)$ for the second line source. Unlike current approaches
focused on cross-correlation functions, the uncertainties in these
linear parameters are fully incorporated into the uncertainties in
any other parameter estimate.

Given these definitions, the probability of the data $\bfy$ given the
linear coefficients $\bfq$, the intrinsic light curves $\bfs$, and
any other parameters of the model $\bfp$ is
\begin{align}
   &P\left(\bfy \bigl| \bfq,\bfs,\bfp \right) \propto \left| S N
   \right|^{-1/2} \nonumber \\
   &\int d^n \bfn \, d^n \bfs \; 
\delta\left(\bfy-\left(\bfs+\bfn+L\bfq\right)\right)
   \exp\left( -{ \bfs^T S^{-1} \bfs + \bfn^T N^{-1} \bfn \over 2} \right).
\end{align}
After evaluating the Dirac delta function, we ``complete the
squares'' in the exponential with respect to both the unknown
intrinsic source variability $\bfs$ and the linear coefficients
$\bfq$. This exercise determines our best estimate for the intrinsic
variability 
\begin{equation}
   \bfhs =  S C^{-1} (\bfy-L\bfhq)
  \label{eqn:shat}
\end{equation}
and the linear coefficients
\begin{equation}
   \bfhq =  (L^T C^{-1} L)^{-1} L^T C^{-1} \bfy \equiv C_q L^T C^{-1} \bfy
\end{equation}
where $C = S+N$ is the overall covariance matrix of the data and
$C_q=(L^T C^{-1} L)^{-1}$. With these definitions we can factor the
argument of the exponential into
\begin{align}
   &P\left(\bfy \bigl| \bfq,\bfs,\bfp \right) \propto \left| S N
   \right|^{-1/2} \nonumber \\
   &\exp\left( -{ \bfds^T (S^{-1}+N^{-1}) \bfds \over 2 } 
   -{ \bfdq^T C_q^{-1} \bfdq \over 2 } -{ \bfy^T C_\perp^{-1} \bfy \over 2 }\right)
\end{align}
where 
\begin{equation} 
    C_\perp^{-1} = C^{-1}-C^{-1}L C_q L^T C^{-1}
\end{equation}
is the component of $C$ that is orthogonal to the fitted linear
functions, the variances in the linear parameters are
\begin{equation}
   \langle \bfdq^2 \rangle = (L^T C^{-1} L)^{-1} \equiv C_q,
\end{equation}
$\bfds=\bfs-\bfhs$ and $\bfdq=\bfq-\bfhq$. We can marginalize the
probability over the light curve $\bfs$ and the linear parameters
$\bfq$ under the assumption of uniform priors for these variables to
find that
\begin{align}
    &P\left(\bfy \bigl| \bfp \right) \propto \mathcal{L} \nonumber \\
    &\equiv \left|S+N\right|^{-1/2}\left|L^T
   C^{-1}L\right|^{-1/2} \exp\left(  -{ \bfy^T C_\perp^{-1} \bfy \over 2 } \right)
    \label{eqn:likefit}
\end{align}
where $\mathcal{L}$ represents the likelihood function we are to
maximize, and the remaining parameters $\bfp$ are those describing the
process (Equation~\ref{eqn:cfunc}) and the transfer functions. The
term in the exponent, $\bfy^T C_\perp^{-1} \bfy$, is the generalized
$\chi^2$ that we present throughout the paper. While this treatment
of linear parameters was included by \cite{RP92}, \cite{RK94} chose
to subtract fixed means rather than marginalizing over them as part
of the analysis as we do here. The variance in the estimate for the
mean light curve is
\begin{equation}
   \langle \bfds^2 \rangle = S - S^T C_\perp S.
   \label{eqn:svar}
\end{equation}

We can estimate the light curve $s(t)$ at any unmeasured time using
the same formalism. The simplest means of doing so is simply to pad
the data vector $\bfy_d$ with additional fake points $\bfy_f$ that
have infinite measurement uncertainties in the sense that
$N^{-1}\rightarrow 0$ for these points. After appropriately
partitioning the matrices, the estimate of the light curve at the
unmeasured points is
\begin{equation}
   {\bfhs}_f = S_{fd} (S_{dd}+N_{dd})^{-1} \bfy_d
    \label{eqn:interp}
\end{equation}
with variance relative to the true light curve of
\begin{equation}
   \langle {\bfds}_f^2 \rangle = S_{ff}-S_{fd}  (S_{dd}+N_{dd})^{-1} S_{df}.
    \label{eqn:varinterp}
\end{equation}
where $S_{dd}$, $S_{ff}$, $S_{fd}$ and $S_{df}$ are the data-data,
fake-fake, fake-data and data-fake covariance matrices of the process
and $N_{dd}$ is the noise matrix of the data. The inclusion of the
fake points has no effect on the expected results for the measured
data points. 

\begin{figure*}[t]
\epsscale{1.0} \plotone{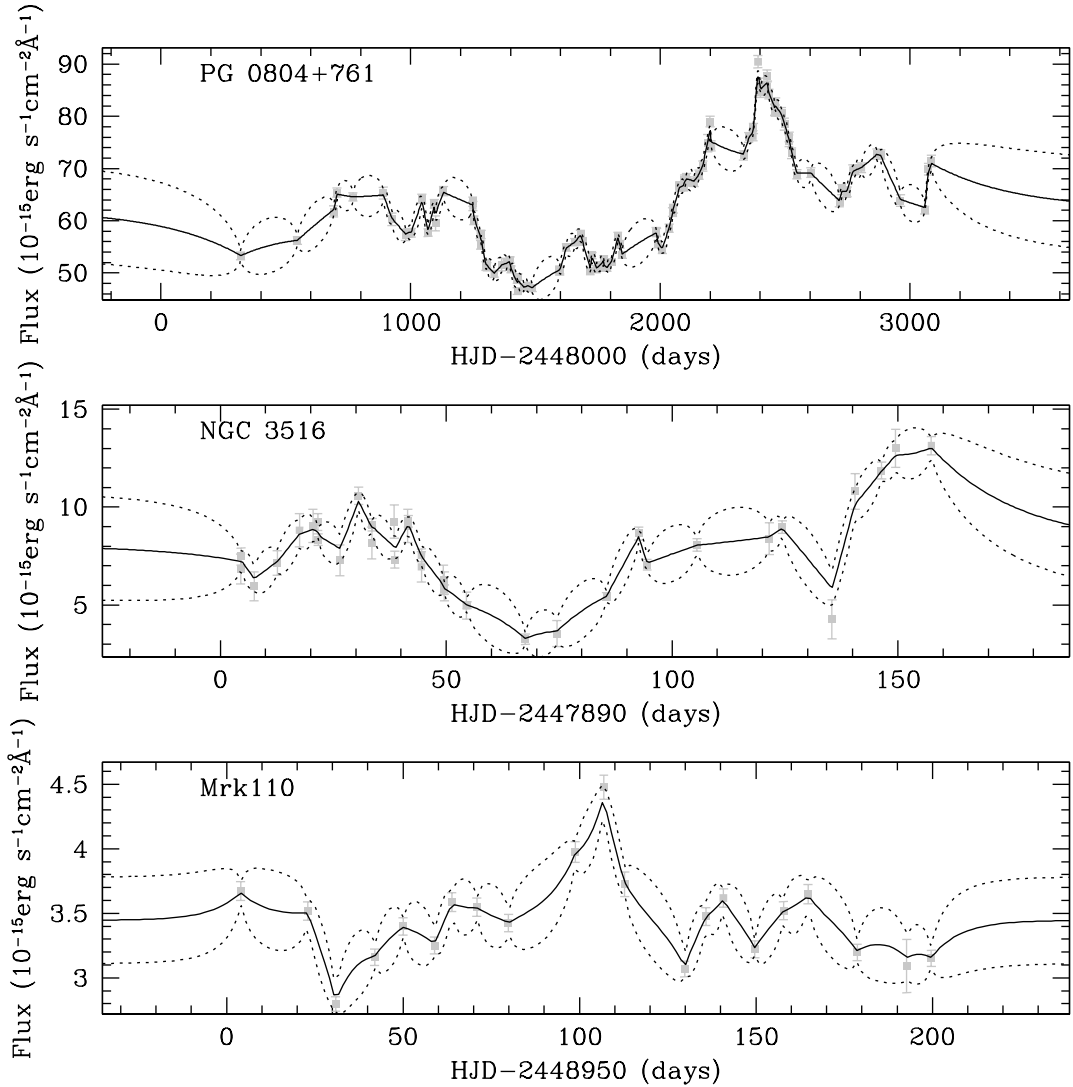}
\caption{Continuum models. The solid line shows the expected mean
source light curve $\bfhs$~(Equation~\ref{eqn:shat}) and the dashed
line shows the expected spread~(Equation~\ref{eqn:svar}) of light
curves about the mean consistent with the data. An individual light
curve realization consistent with the data~(see
Equation~\ref{eqn:fastpredict}) will show more structure than this
mean light curve and have excursions outside the dashed line
consistent with the estimated variance.} 
\label{fig:contlc} 
\end{figure*}

Just to re-emphasize the point, this formalism was first outlined by
\cite{RK94} based on \cite{PRH92} and \cite{RP92}. We have refined it
slightly to use a specific process model, to optimize the parameters
of that model and to include the means of the light curves as
parameters that are automatically marginalized. Unfortunately, we
will not be able to use the fast implementation of this method for
exponential covariance matrices from \cite{RP95}, because the
inclusion of the transfer functions means that $S$ is not a simple
exponential covariance matrix and hence does not have a simple,
tridiagonal inverse for the fast method.

We can, however, use the fast methods for generating simulated light
curves.  In particular, we are interested in light curves constrained
to resemble the continuum light curve. As discussed by \cite{RP92},
such a light curve is simply the estimated mean light curve given by
Equation (\ref{eqn:shat}) with an added random component that has the
covariance matrix $Q = (S^{-1}+N^{-1})^{-1}$. \cite{RP92} suggest
determining the eigenmodes of $Q$ which are then the independent
``normal'' modes that can be added to the mean light curve to produce
a random realization constrained by the continuum light curve. This
is computationally expensive. Instead, we note that if we Cholesky
decompose $Q = M^T M$, where $M$ is an upper triangular matrix, and
define the random component of the light curve by ${\bf u} = M {\bf
r}$ where ${\bf r}$ is a vector of zero-mean, unit-dispersion
Gaussian random deviates, that 
\begin{equation}
 \langle {\bf u} {\bf u}^T \rangle = M \langle {\bf r} {\bf r}^T \rangle M^T
 = M M^T = Q^T = Q
 \label{eqn:fastpredict}
\end{equation}
since the covariance matrix $\langle {\bf r} {\bf r}^T \rangle$ of
the Gaussian deviates is simply the identity matrix and $Q$ is
symmetric. Since $Q^{-1}$ is a tridiagonal matrix given the
exponential covariance matrix and a diagonal noise matrix, we can
generate very high dimension ${\bf u}$ that can be convolved with the
transfer function to produce a simulated line light curve in
$\mathcal{O}(K)$ operations rather than the $\mathcal{O}(K^3)$ needed
following the eigenmode approach.

The original application of the method by \cite{PRH92} was to cross
correlate the light curves of two images of a lensed quasar in order
to estimate the time delay between them. While this was not discussed
in terms of transfer functions, it does correspond to a transfer
function of the form $\Psi(t_i-t_j)=\delta(t_i-(t_j+\Delta t))$,
making the second light curve a lagged version of the first.
\cite{PRH92} also treated the parameters corresponding the process as
fixed parameters, derived by fitting a power law to the structure
functions of the light curve. It is likely that some combination of
neglecting uncertainties in the process model or covariances in the
errors of the light curves led \cite{PRH92} to obtain an incorrect
estimate of the time delay despite the elegance of the approach.

In Rybicki \& Kleyna's~(1994) expansion of the method to
reverberation mapping, they used rising and falling sawtooth and
isosceles triangle transfer functions, finding little difference
between the results or ability to discriminate between them. Thus,
for this initial reconnaissance, we will simply use a top-hat
(rectangular function) for the transfer function, 
\begin{equation}
    \Psi(t-t') = A \left(t_2-t_1\right)^{-1} \quad\hbox{for}\quad t_1 \leq t-t'
    \leq t_2
    \label{eqn:tophat}
\end{equation}
which has a mean lag of $\langle \tau \rangle=(t_1+t_2)/2$ and a
temporal width of $\Delta \tau = t_2-t_1$. The necessary integrals
for Equations~(\ref{eqn:lc}), (\ref{eqn:lauto}), and
(\ref{eqn:lcross}) are all analytic (see the Appendix) and the model
includes the limits of a delta function as $\Delta \tau \rightarrow
0$ and a uniform thin shell as $t_1 \rightarrow 0$.  The scaling
coefficient $A$ determines the line response for a given change in
the continuum~(i.e., the responsivity of BLR clouds), but for present
purposes we will largely view it as a nuisance variable.

We use the {\it amoeba} minimization method \citep{nr} to optimize
the solution and then either a Monte Carlo Markov Chain~(MCMC,
\citealt{MRRTT53, hastings70}) or optimization over a grid to
estimate parameter uncertainties. We carry out the analysis in two
phases. We first analyze the continuum light curve on its own, using
logarithmic priors for $\tau_{\rm d}$ and $\hat{\sigma}$ to determine
the range of the variability process parameters consistent with the
continuum light curve. The logarithmic prior on $\tau_{\rm d}$
essentially penalizes values that deviate from the median sampling
intervals to avoid both unphysically large $\tau_{\rm d}$ and a
second class of solutions of $\tau_{\rm d} \rightarrow 0$, when all
data are completely uncorrelated and the model simply uses $\sigma$
to broaden the uncertainties until obtaining an acceptable fit. Then
we do the joint analysis of the continuum and the lines using
Gaussian priors for $\tau_{\rm d}$ and $\hat{\sigma}$ determined from
the analysis of the continuum in isolation. In detail, we take the
results of the MCMC analysis of the continuum and used uncorrelated
priors on $\ln\tau_{\rm d}$ and $\ln\hat{\sigma}$ (which is
conservative), where the prior for each variable was centered at the
median value with the Gaussian width chosen to match the upper and
lower $1\sigma$ confidence regions. We then used uniform priors for
$A$, $t_1$ and $t_2$. 

\begin{figure}[t]
\epsscale{1.0} \plotone{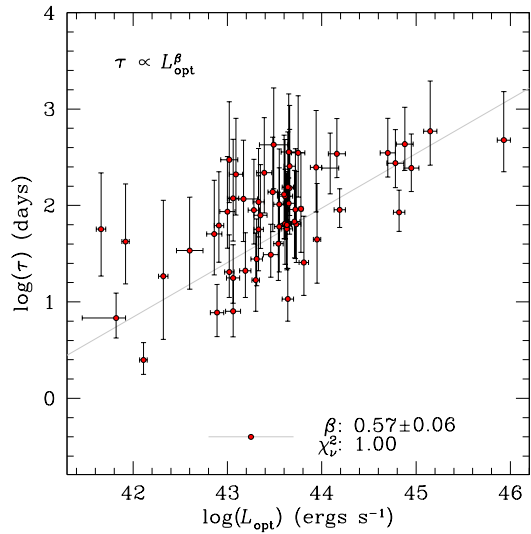}
\caption{Rest-frame damping timescale $\tau_{\rm d}$ of the continuum
light curves as a function of optical luminosity. The uncertainties
in $\tau_{\rm d}$ are the $\pm 1\sigma$ range.}
\label{fig:taul}
\end{figure}

The reason for using the continuum to define a stronger Gaussian
prior on the process variable before carrying out the joint analysis
is to eliminate the aforementioned second class of solutions of
$\tau_{\rm d} \rightarrow 0$ that could potentially bias our lag
estimates. This secondary solution always exists at some level
because of the finite temporal sampling. For modeling the continuums,
we are only analyzing cases with significant variability, so this is
not an issue for the individual light curves. However, in the joint
analysis, if we fit the line and continuum light curves
simultaneously at the wrong lag, the optimal solution will be to let
$\tau_{\rm d} \rightarrow 0$ since there are then no correlations
between data points.  Physically, it made more sense to consider only
the ranges for the process variables $\tau_{\rm d}$ and
$\hat{\sigma}$ that were statistically consistent with the continuum
variability.


\section{The Statistical Process Model of the Continuum}
\label{sec:cont}
This approach depends on using a statistical model for the
variability process of the continuum in order to optimally model the
underlying light curve of the continuum. Here we use the exponential
covariance matrix suggested by \cite{kelly09}, although it was also
introduced by \cite{RP95} to enable a fast version of the SPEAR
approach. Physically, the exponential covariance matrix in
Equation~\ref{eqn:cfunc}
corresponds to a damped random walk with an amplitude scale
$\hat{\sigma}$ and a damping time scale $\tau_{\rm d}$. On long time
scales the variance of the light curve is $\hat{\sigma}(\tau_{\rm
d}/2)^{1/2}$ and on short time scales it is $\hat{\sigma}\sqrt{t}$.

\begin{figure*}[t] 
\epsscale{1.0} \plotone{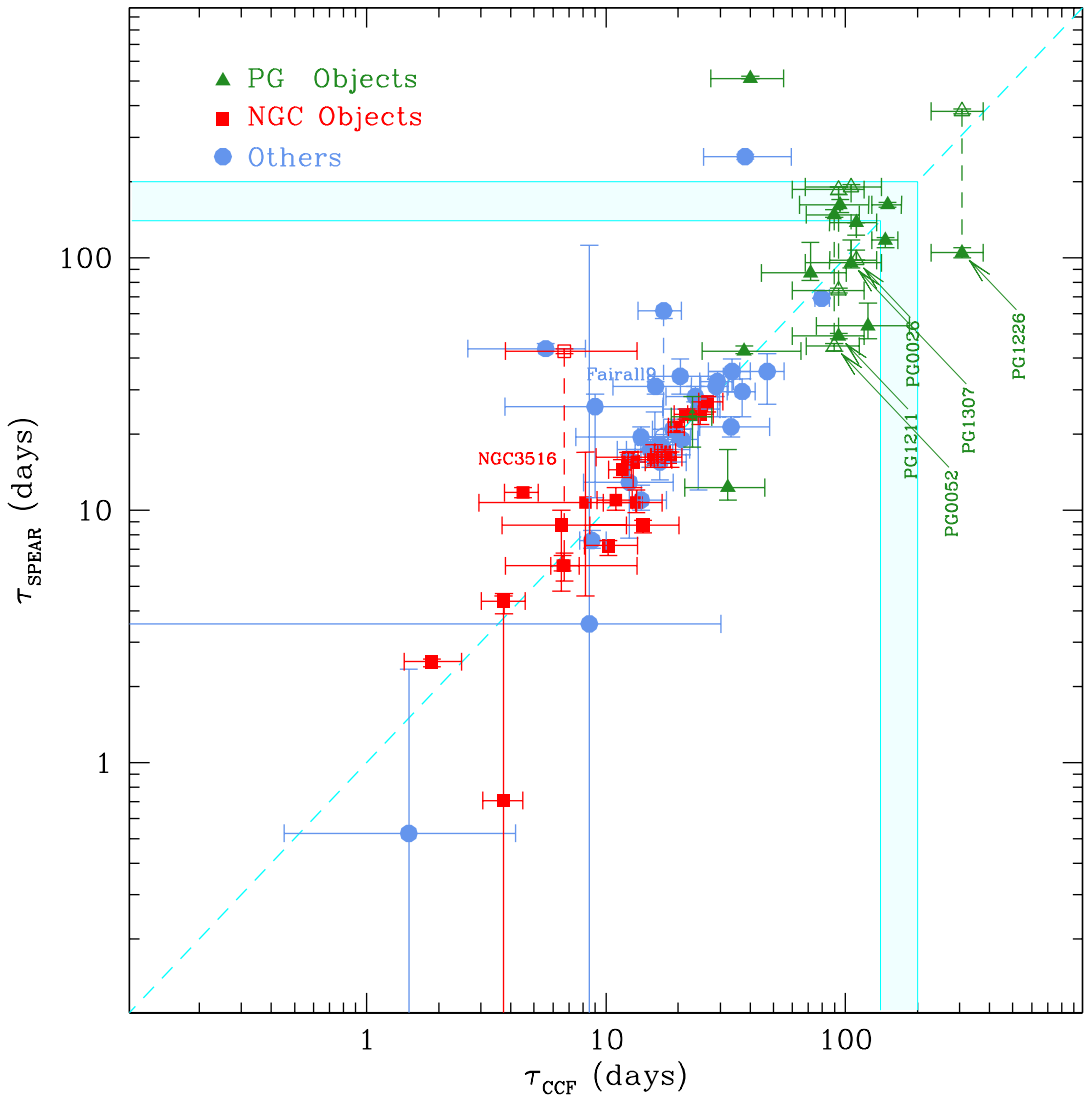}
\caption{Comparison of rest-frame H$\beta$ time lags from the CCF and
the SPEAR methods. Green triangles, red squares and blue circles were
used for the PG, NGC and other objects, respectively, with $\pm 1
\sigma$ error bars indicated on both estimates. The labeled points
linked by dashed vertical lines are objects having multiple lag
solutions and the filled symbol is the higher likelihood solution.
The two intersecting stripes indicate the region where the solutions
from {\it both} methods may be false due to the seasonal
gap~(140--200 days, with time dilation).}
\label{fig:hbetacomp}
\end{figure*}

\cite{kelly09} use this to model the light curves of 100 quasars,
including some of the objects we will consider here, using a light
curve forecasting approach to estimate the process parameters.
\cite{koz10a} show how the \cite{kelly09} approach can be derived
from the SPEAR approach and demonstrated that forecasting is less
statistically optimal for parameter estimation than using the
complete light curve modeling method of SPEAR, and then applied the
process model and the SPEAR method to the OGLE-III \citep{udalski08}
light curves of $\sim 2500$ mid-infrared-selected quasars behind the
Magellanic Clouds \citep{KK09}. They confirm that the damped
random-walk model describes quasar light curves well, and that
quasars occupy a well-defined region of $\tau_{\rm
d}$--$\hat{\sigma}$ parameter space. This is further confirmed by
\cite{macleod10}, who used this approach to model 9,000 SDSS quasars
to examine the correlations of $\sigma$ and $\tau_{\rm d}$ with other
quasar properties.

Unlike the previous papers, we fit flux rather than magnitude light
curves because the line flux is more closely related to the continuum
flux than to the continuum magnitude. Thus, we start by examining how
well the damped random-walk process models the 60 continuum flux
light curves for the 31 systems we consider in \S\ref{sec:hbeta}.
Figure~\ref{fig:chi2dof} shows the distribution of the $\chi^2$ per
degree of freedom for the best-fit models of all the continuum light
curves we consider. Since half of the continuum light curves in our
sample have less than 50 data points, the expected $\chi^2/dof$
distribution is broader than that of the OGLE light curves~($\sim$
500 points) considered by \cite{koz10a}.  Nevertheless, the
$\chi^2/dof$ distribution indicates that the statistical process
model provides a reasonable fit to the light curves.  The fact that
the distribution is narrower than expected for correctly estimated
Gaussian uncertainties suggests that the reported photometric errors
are somewhat larger than the true uncertainties, or that there has
been some pruning of outliers from the light curves.

\begin{figure*}[t]
\epsscale{1.0} \plotone{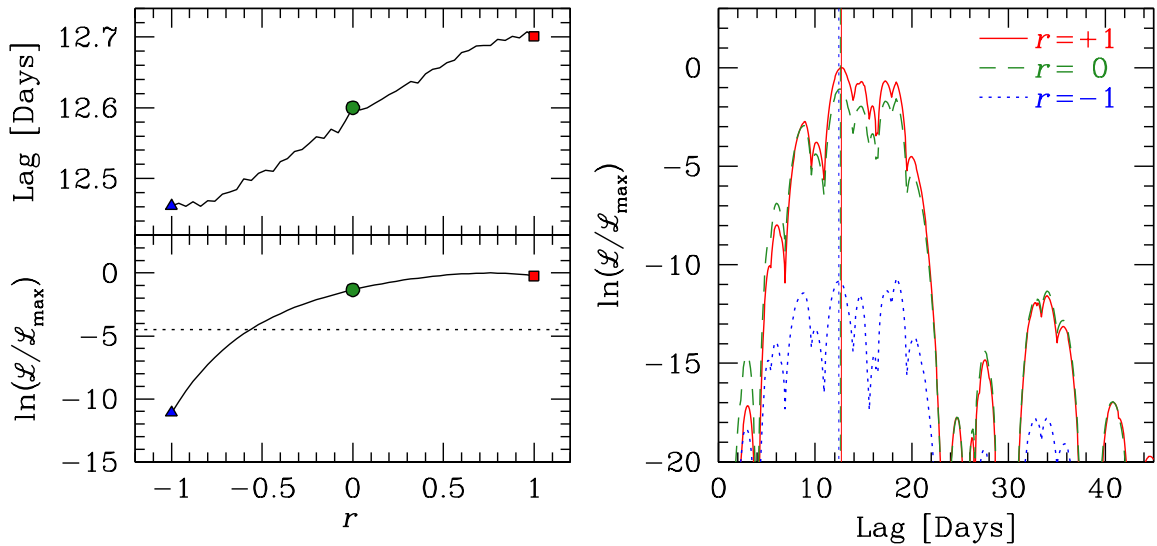}
\caption{Sensitivity of the lag estimates to the noise correlation
coefficient $r$ between the H$\beta$ and the continuum light curves
of PG~0844. The left top panel shows the dependence of the lag on the
correlation coefficient $r$. The left bottom panel shows the
corresponding change in the likelihood function with $r$ at the
best-fit lag. In these panels, the blue triangle, green circle, and
red square mark the results for $r= -1,0,+1$, respectively, and the
dotted line indicates the $3\sigma$ limit of the likelihood function.
The right panel compares the lag likelihood distribution for these 3
cases: $r=-1$~(blue dotted curve), $0$~(green dashed), and $+1$~(red
solid), respectively. The dashed lines in the two right panels
indicate the position of the best-fit lag, which is almost the same
for all 3 cases.}
\label{fig:covtest} 
\end{figure*}

Figure~\ref{fig:contlc} shows three examples of modeled continuum
light curves interpolated and extrapolated from
Equation~(\ref{eqn:shat}) and their uncertainties from
Equation~(\ref{eqn:svar}), as well as the observed light curve.  The
estimated light curve at time $t$ is in essence a weighted average
over data points within the damping time $|t-t'| \ltorder \tau_{\rm
d}$ that balances the variance expected on those time scales due to
the process against the uncertainties in the data point to determine
how closely the model light curve approaches a particular data point.
Far from any data points, the model returns to the light curve mean
on the time scale $\tau_{\rm d}$.  Remember, however, that
Equation~(\ref{eqn:shat}) is an estimate for the average of all
possible light curves that could be drawn from the process that would
be consistent with the data --- a particular realization of such a
light curve would show additional structure (see \citealt{RP92}). The
``error snake'' surrounding the model light curve is the variance in
these possible light curves. Near data points, its width approaches
that of the measurement errors and then grows as the distance $\Delta
t$ from any data point increases. The variance from the process
initially increases as $\hat{\sigma}|\Delta t|^{1/2}$, but then
saturates at the overall process variance once $|\Delta t| \gg
\tau_{\rm d}$. Thus, in the extrapolated regions we see the model
light curve becomes a constant and the error snake expands and then
becomes constant. 

The three objects shown in Figure~\ref{fig:contlc} represent three
typical levels of light-curve sampling quality for the objects we
consider. Generally, the light curves of the Palomar--Green~(PG)
quasars obtained by \cite{kaspi00} were sampled every 1--4 months
over a baseline as long as 7.5 yr, as opposed to most of the
low-luminosity Seyfert 1 AGNs that were more densely sampled over
shorter baselines. The rest of the sample mainly consists of nearby
bright Seyfert galaxies \citep{peterson98} whose light curves are
sparsely sampled over a short baseline. \cite{peterson04} discuss the
data in detail. In addition, we also include new light curves from a
recent high sampling rate, multi-month reverberation mapping campaign
on six local Seyfert galaxies~\citep{denney10}. 

We expect the damping timescales $\tau_{\rm d}$ to show correlations
with the physical characteristics of the accretion disk such as the
mass of the central black hole, and the AGN luminosity
\citep{peterson08}.  \cite{kelly09} demonstrated this scaling
relationship between $\tau_{\rm d}$ and $L_{\rm AGN}$ by performing a
linear regression of $\tau_{\rm d}$ on $L_{\rm AGN}$, while
\cite{CB01} also found a positive correlation between the
characteristic timescale and black hole masses. Their characteristic
timescale, which is defined by the timescale where the structure
functions flattened, is roughly equivalent to $\tau_{\rm d}$.
Figure~\ref{fig:taul} shows that the more luminous central engines
have longer exponential damping timescales, as we would expect from
\cite{kelly09}, up to any minor differences from fitting fluxes
rather than magnitudes. Note that \cite{macleod10} argue that the
dependence of $\tau_{\rm d}$ on black hole mass $M_{\rm BH}$ is the
real driver of the correlation between $\tau_{\rm d}$ and luminosity.
We can use these correlations to estimate $\tau_{\rm d}$ for sources
lacking sufficiently good light curves. 

\begin{figure*}[t]
\epsscale{1.0} \plotone{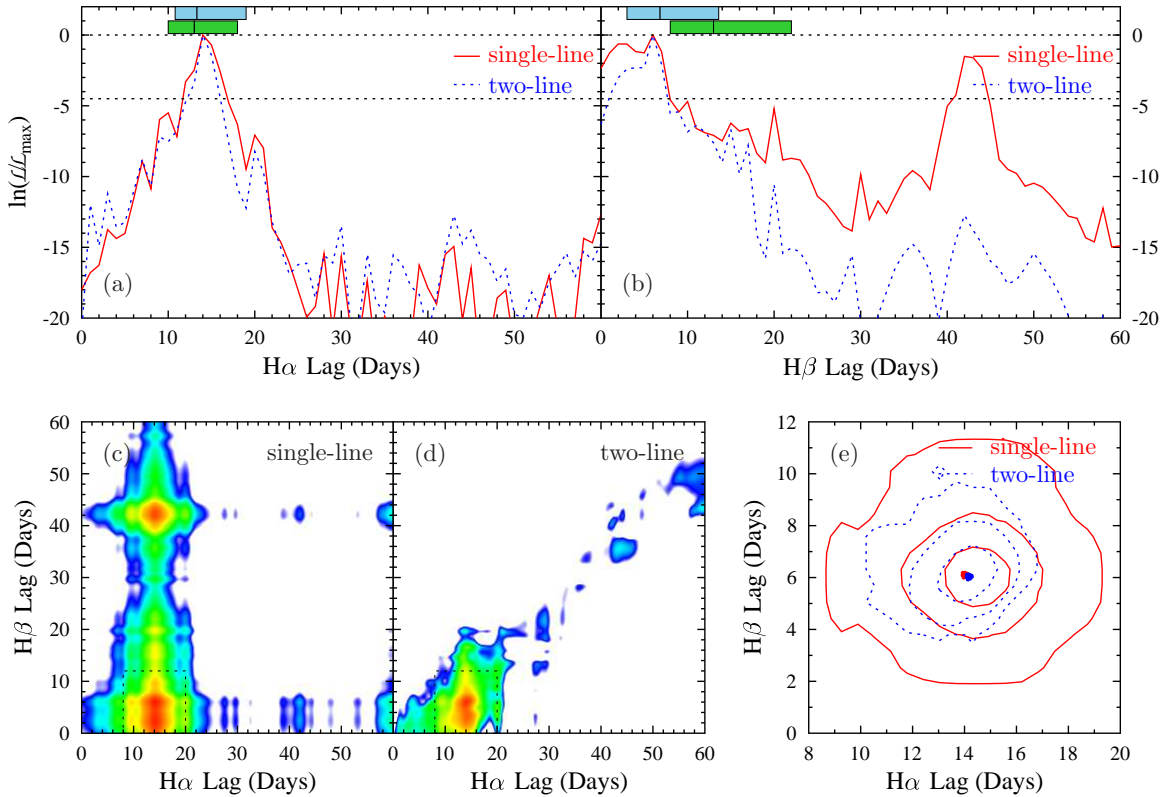}
\caption{Comparison between independent~(single-line) and
joint~(two-line) fits to the H$\alpha$ and H$\beta$ light curves of
NGC~3516. The red solid lines are the estimate from the single-line
fits, while the blue dashed lines are those from the two-line fits.
The top left~(right) panel compares the likelihood distributions of
the two fits for the H$\alpha$~(H$\beta$) line. The interval between
the two dotted lines corresponds to a $3\sigma$ range in the
likelihood, while the two blocks above indicate the $\pm 1\sigma$
range of the CCF peak analysis~(upper) and CC centroid
distribution~(lower), where the central lines mark the $\tau_{peak}$
and $\tau_{cent}$ values, respectively. The two bottom left panels
shows the color-coded covariance map between the two lags for the
single-line and two-line fits, respectively. The contours in the
bottom right panel compare the confidence levels calculated from MCMC
for the H$\alpha$/H$\beta$ lags near the peak~(black boxes inside
left two panels). Working outward, the three contour curves are for
$1\sigma$, $2\sigma$ and $3\sigma$ levels, respectively. Note that
those are all observed-frame lags and the H$\beta$ light curve here
is the older of the two we have for NGC~3516.}
\label{fig:joint2_ngc3516}
\end{figure*}

%
\section{Estimating Emission-Line Lags} 
\label{sec:hbeta}

As our first application of the SPEAR method we recompute the lags of
101 emission-line light curves for 31 objects in the literature~(the
compilation of \citealt{peterson04} with the addition of data from
\citealt{bentz06b}, \citealt{grier08}, and \citealt{denney06,
denney10}). We carried out the analysis in three stages. First, as
discussed in \S\ref{sec:cont}, we modeled the continuum alone to
determine the range of process parameters ($\tau_{\rm d}$,
$\hat{\sigma}$) consistent with the continuum light curve. We use
this distribution of estimated $\tau_{\rm d}$ and $\hat{\sigma}$ as a
prior for the joint models of the continuum and line light curves in
order to avoid the secondary solutions with $\tau_{\rm d} \rightarrow
0$ as discussed in \S\ref{sec:method}. Second, for each joint model,
we find the best-fit top hat transfer
function~(Equation~\ref{eqn:tophat}) which maximizes the model
likelihood calculated by Equation (\ref{eqn:likefit}), along with an
updated set of process parameters. Finally, we ran an MCMC analysis
on each joint model to calculate the statistical confidence limits on
each best-fit parameter found by global optimization on a grid,
especially the time lag. We then compare these estimates to those
derived from previous CCF analyses. We refer to these models as the
``single-line'' fits since they are solving for a single top-hat
transfer function. The dotted histogram in Figure~\ref{fig:chi2dof}
shows the $\chi^2/dof$ distribution of the single-line model. It has
a similar shape to the $\chi^2/dof$ distribution of the stochastic
model for only the continuum light curve, and confirms that the
statistical model provides a good fit to the quasar variability, as
well as the overall {\it ansatz} that the H$\beta$ variability is a
scaled and smoothed version of the continuum. The $\chi^2/dof$
distribution of the single-line model is somewhat worse than for
fitting the continuum alone, but still reasonably consistent with
statistical expectations.

For the sake of uniformity of emission-line species in the comparison
between the SPEAR and the CCF methods, and to avoid confusion in the
figures for sources with multiple line observations, we will focus on
the the 66 H$\beta$ light curves in our subsequent analyses, and
tabulate all the other emission-line lags we successfully computed
with SPEAR in Table~\ref{tab:lags}.

\begin{figure*}[t]
\epsscale{1.0} \plotone{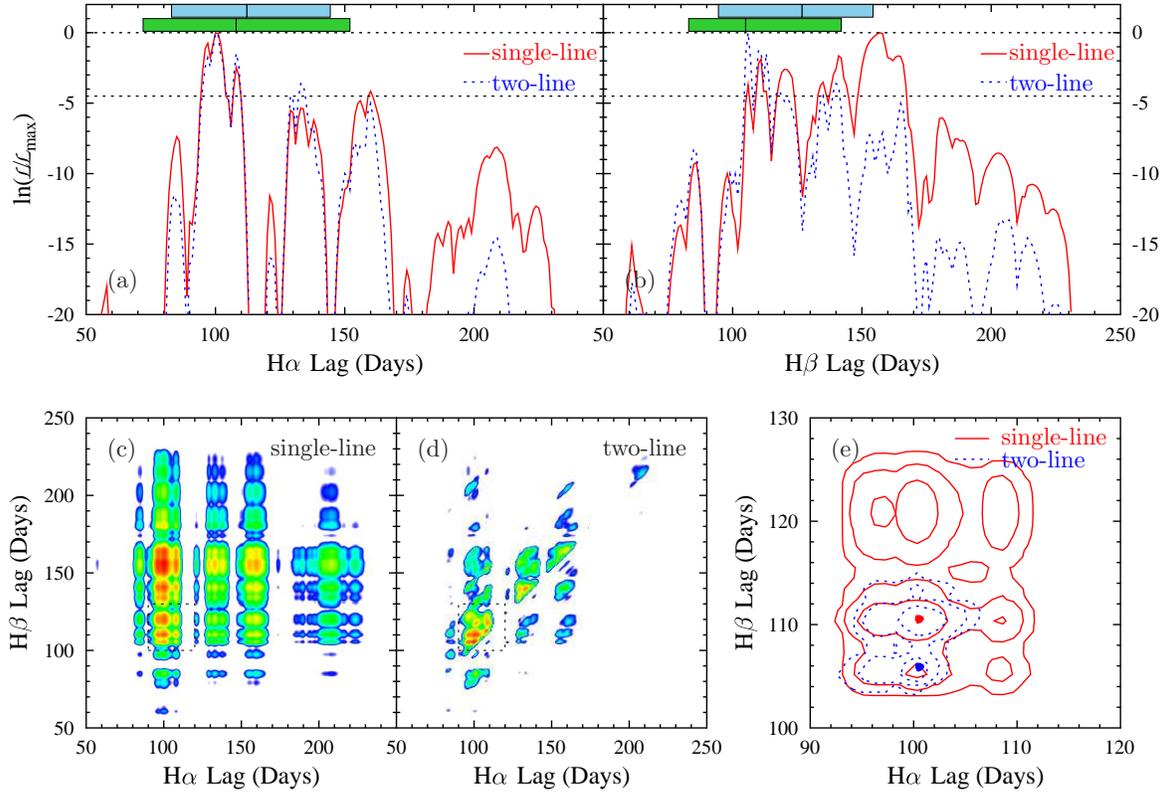}
\caption{Comparison between independent~(single-line) and
joint~(two-line) fits to the H$\alpha$ and H$\beta$ light curves of
PG~0026. The format is the same as in
Figure~\ref{fig:joint2_ngc3516}.}
\label{fig:joint2_pg0026}
\end{figure*}

Figure~\ref{fig:hbetacomp} shows the comparison between CCF centroid
time lags $\tau_{\rm CCF}$ and our lags $\tau_{\rm SPEAR}$ for all
the H$\beta$ lines. The range of uncertainties for $\tau_{\rm CCF}$
contains $68.3\%$ of Monte Carlo realizations in the
cross-correlation centroid distribution~(CCCD), while our estimated
error boundaries are defined by the $68.3\%$~(${\rm
ln}\mathcal{L}/\mathcal{L}_{\rm max}=0.5$) confidence levels that
encloses the best-fit lags (i.e., $\pm 1 \sigma$ errors if the
probability distribution is Gaussian in both cases). Based on the
structure of the lag probability distribution, we can divide the
``single-line'' fits into five quality groups:
\begin{itemize}
    \item[(\Rmnum{1})] In most of the cases~(43 of 66), the
        likelihood distribution for the lags has a single peak and
        there is an unambiguous H$\beta$ lag.
    \item[(\Rmnum{2})] In 9 cases, the likelihood distribution has
        multiple peaks with significant ($> 3\sigma$) likelihood
        differences.  This occurred for one season of Akn~120
        (JD49980--50175)\footnote{For brevity, we retain only the
        five least significant digits of the Julian Date.}, Mrk~110
        (JD48953--49149), and Mrk~590 (JD49183--49338); two seasons
        of Mrk~79 (JD48193--48393 and JD49996--50220); NGC~4051,
        PG~0844\footnote{For brevity, we truncate the PG coordinate
        identifiers to right ascension only since this introduces no
        ambiguity in the present small sample.}, PG~1411, and
        PG~1617. Compared to our estimate, the CCF analysis picks a
        lower likelihood peak or aliases several peaks into one broad
        peak. Generally, the two peaks are so close that the
        differences between the results from the two methods are
        insignificant compared to the uncertainties. 
    \item[(\Rmnum{3})] In 7 cases, the likelihood distribution has
        multiple peaks of comparable significance~($\le 3\sigma$: one
        series of NGC~3516 (JD47894--48047), Fairall~9, PG~0026,
        PG~0052, PG~1211, PG~1226, and PG~1307).  They are shown in
        Figure~\ref{fig:hbetacomp} as the objects with a dashed line
        connecting the possible solutions. The traditional CCF method
        seems to find one broad peak for these sources, rather than
        multiple peaks, leading to large reported uncertainties for
        the estimate of $\tau_{\rm CCF}$.  These degeneracies are
        largely caused by poor light curve sampling that allows the
        light curve of the emission line to be mapped into the
        sampling gaps of the continuum.  This problem is worst for
        the PG objects, which have many ``seasonal gaps'' over the
        long observing baselines ($\sim$ 7.5 yr), leading to a
        clustering of solutions around 180 days in the
        observed-frame. Such seasonal aliasing problems affect the
        CCF-based methods as well~\citep{grier08}. 
    \item[(\Rmnum{4})] In four cases, the light curves are very
        poorly sampled: IC~4329A, one season on NGC~4593
        (JD47894--48049), one season of Mrk~279 (JD47205--47360), and
        one season of NGC~3227 (JD47894--48045). These cases were
        also flagged as unreliable by \cite{peterson04}, so we
        exclude them from our subsequent analyses.
    \item[(\Rmnum{5})] The lags derived from the SPEAR method appear
        to be wrong in two cases, 3C\,120 and PG~1613. We also
        exclude both from our subsequent analyses. The 3C\,120 light
        curves have a baseline of 7 years, but are very sparsely
        sampled. The CCF method finds a lag $\sim$ 40 days in the
        observed frame. Although we find a sub-peak at 40 days, the
        model favors another peak of much higher significance at 259
        days. For PG~1613 we obtain a lag of $\sim$ 575 days in the
        observed-frame, much larger than the $\sim$ 50 day CCF
        estimate. In both cases, the longer lag is favored because it
        minimizes the data overlap --- 259 and 575 days put most of
        the line data in the seasonal gaps, and many points also lie
        before the start of the continuum light curve.  This is
        essentially an aliasing problem in our method. We also note
        that the continuum flux varied by up to 50\% over the 7 year
        span of the light curves. We know empirically that the
        scaling coefficient $A$ in the transfer function is inversely
        correlated with ionizing continuum flux~(see the right panel
        of Figure~\ref{fig:lrs} and the discussion in
        \S\ref{sec:breath}), but we treat $A$ as a constant parameter
        in each individual fit.  This may create problems for light
        curves with the significant long term trends observed for
        these objects. Allowing $A$ to vary and adopting a prior that
        penalizes large lags that minimize light curve overlap would
        likely solve these problems.
\end{itemize}
%
\section{Model Test for Correlated Errors} 
\label{sec:corr}

Correlated errors have long been viewed as a problem in traditional
CCF analysis. Observations made at a common epoch are inevitably
correlated by the processes required for calibration, light curve
extraction, broad/narrow line modeling and removal of host or Fe{\sc
ii} contamination. Because no assumption about the properties of the
noise matrix $N$ was made in \S\ref{sec:method}, it is easy to
include the effects of correlated errors within our approach. While
we did not make an extensive survey of our ability to model noise
correlations between the continuum and lines, we did carry out some
experiments for objects noted as potentially having strong
covariances by \cite{peterson04}. 

The simplest test is to introduce a covariance factor $-1 \leq r \leq
1$ and add off-diagonal terms to the noise matrix $N$ for line and
continuum points measured at the same epoch of $N_{cl}(t,t) =
N_{lc}(t,t) = r \sigma_l(t) \sigma_c(t)$ in order to examine the
sensitivity of the lag estimates to correlated noise between the line
and the continuum measured at each epoch.  This should be present in
the data at some level because of the challenge of consistently
subtracting the contribution of the host galaxy to the line and the
continuum in the presence of variable observational conditions.

\begin{figure*}[t]
\epsscale{1.0} \plotone{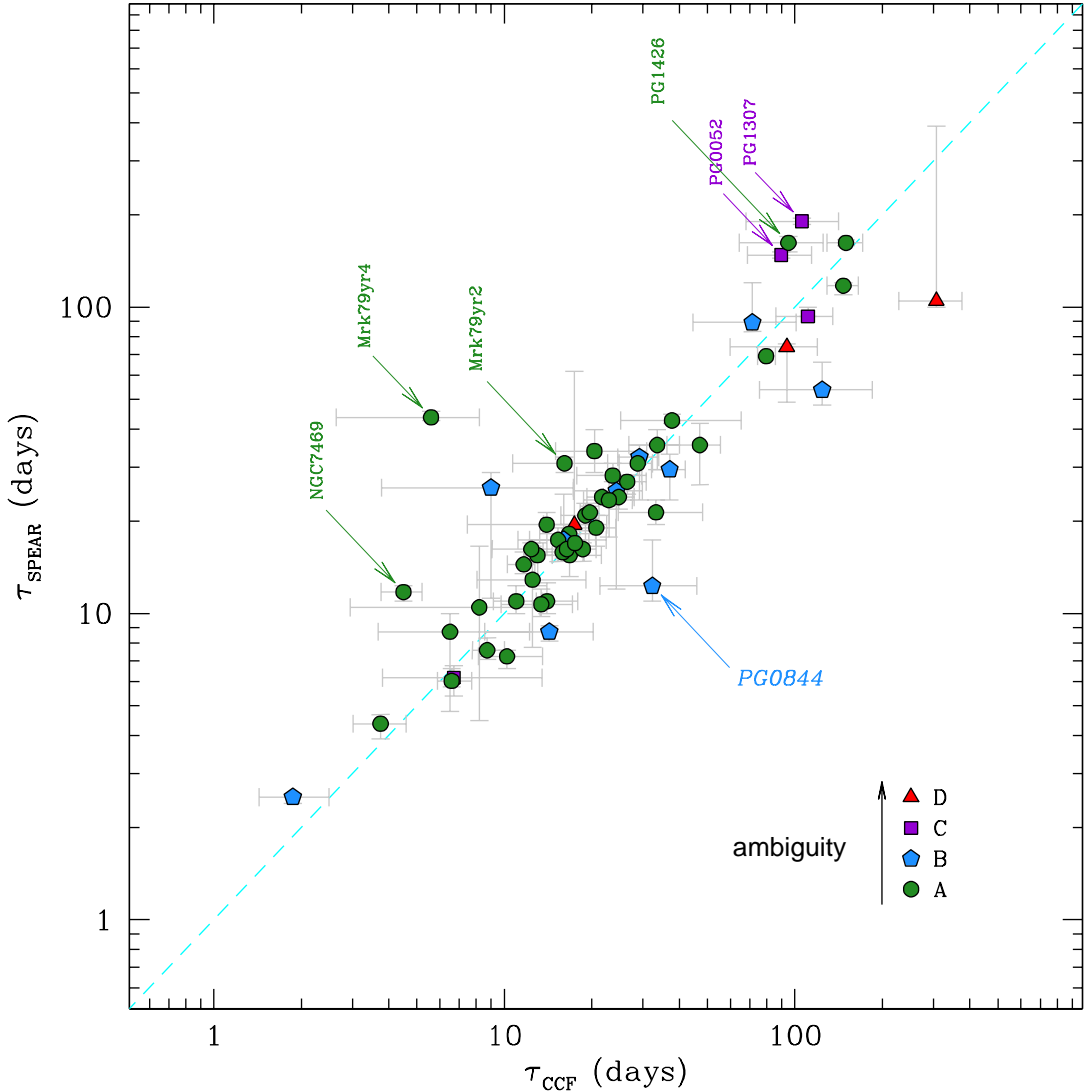}
\caption{Comparison of the H$\beta$ time lags from CCF analysis and
the SPEAR method, similar to Figure~\ref{fig:hbetacomp}, updated
where we have used the two-line fits and dropping the 6 unreliable
sources. Four types of symbols are used to indicate our estimate for
increasing levels of ambiguity in the lag estimate. Objects with
inconsistent lag estimates between the two methods are labeled.}
\label{fig:hbetacomp2} 
\end{figure*} 

Figure~\ref{fig:covtest} illustrates the effects of adding
off-diagonal correlated noise terms on the H$\beta$ lag estimate of
PG~0844. The shift in the estimated lag~(left top panel) induced by
$r$ varying from $-1$ to $+1$ is only about 0.25 days, much smaller
than the median sampling interval of the light curves. The
corresponding change in the likelihood~(left bottom panel) shows a
plateau at $r>0$ and slowly asymptotes to a maximum at $r=+1$,
suggesting that the errors in the two light curves are positively
correlated.  The lag likelihood distribution~(right panel) changes if
we assume different levels of correlations $r$ between the light
curves.  While the overall lag likelihood is greatly depressed in the
$r=-1$ case, the likelihood distributions are nearly identical in the
$r=0$ and $r=+1$ cases.  However, the peaks near the best lag
estimate~($\sim$ 12 days) are slightly more significant in the $r=+1$
case than in the $r=0$ case. We explored this issue for several other
systems, and generally the impact on the estimated lag is negligible,
although different levels of (anti-)correlations are detected.


\section{Joint Analysis of Multiple Lines} \label{sec:doublehat}
In \S\ref{sec:hbeta}, we found that poor light curve sampling was a
significant problem in many systems, particularly in objects with
observed-frame lags on time scales similar to the seasonal gap
spacing.  However, if multiple lines have been measured, then we have
significant, additional data to better sample the light curves under
our overall {\it ansatz} that all light curves are scaled, smoothed,
and displaced versions of the continuum. Simultaneous fits also
determine the covariance between the lags of the different lines. In
this section, we explore simultaneously fitting the continuum and two
emission-line light curves~(hereafter ``two-line'' fits, as opposed
to the ``single-line'' fits in \S\ref{sec:hbeta}, as we are now
fitting two top-hat transfer functions).

Figure~\ref{fig:joint2_ngc3516} summarizes the significant
improvement in estimating the H$\beta$ time lag of NGC~3516 of the
\cite{wanders93} data (JD47894--48047) after including the H$\alpha$
light curve~(two-line) compared to using the H$\beta$ line
alone~(single-line).  NGC~3516 is a case where the single-line
H$\beta$ fits shows a secondary peak at $\sim$ 42 days whose
likelihood relative to the main peak at $\sim$ 6 days is high, ${\rm
ln}(\mathcal{L}_{2nd}/\mathcal{L}_{max}) = -1.5$~(solid curve in
panel b). The H$\alpha$ fit does not show such a secondary
peak~(panel a). When we fit both simultaneously, the H$\alpha$ light
curve together with its well-determined lag adds extra information to
the continuum light curve, and thus better constrains the H$\beta$
lag. The second H$\beta$ peak is suppressed and there is a single
unambiguous H$\beta$ lag for the two-line fit~(dotted curve in panel
b). The improvement is most clearly seen in the structure of the
H$\alpha$/H$\beta$ lag likelihood plane~(panel c and d). If we zoom
in on the remaining peak and run a MCMC chain using a flat prior on
lags in the zoomed region, we can see that the two-line fits not only
suppress the secondary peaks but also shrink the uncertainties in the
primary peak to produce better results for both lines~(panel e).

The joint analysis of multiple lines is especially useful for the PG
objects, whose light curves show observational gaps of period
$\sim$180 days in the observed-frame. In the single line fits, the
model would always show (sub)peaks for lags $\sim$180 days because of
the seasonal aliases~(the seasonal stripes in
Figure~\ref{fig:hbetacomp}). It is not possible, however, to do this
for 2 lines simultaneously, so the two-line fits largely eliminate
seasonal aliasing. Figure~\ref{fig:joint2_pg0026} illustrates this
for PG~0026. In particular, the broad H$\beta$ likelihood
distribution shrinks significantly and the maximum likelihood lag
drops from $\sim$160 days to $\sim$106 days ($\sim$ 140 to $\sim$ 93
in the rest frame) and is in better agreement with the H$\alpha$
results.  Although the traditional CCF method makes similar
estimates~(green and blue bands in two top panels), it yields
significantly larger uncertainties by aliasing several peaks into one
broad CCF centroid distribution.
 
\tabletypesize{\scriptsize}
\begin{deluxetable}{lcccc}
\tablecolumns{5}
\tablewidth{0pt}
\tablecaption{Rest-Frame Lag Estimates}
\tablehead{
\colhead{ }&
\colhead{ }&
\colhead{Julian Dates}&
\colhead{   $\tau_{\rm SPEAR}$   }&
\colhead{ }\\
\colhead{  Object  }&
\colhead{Line}&
\colhead{($-$2400000)}&
\colhead{(days)}&
\colhead{Group}\\
\colhead{(1)}&
\colhead{(2)}&
\colhead{(3)}&
\colhead{(4)}&
\colhead{(5)}
}
\startdata
3C 390.3&H$\beta$&49718--50012&$27.9^{+2.4}_{-1.5}$&A\\
3C 390.3&Ly$\alpha$&49718--50147&$11.9^{+34.5}_{-4.6}$&D\\
3C 390.3&C\,{\sc iv}\,$\lambda1549$&49718--50147&$15.0^{+2.0}_{-3.0}$&C\\
Akn 120&H$\beta$&48148--48344&$35.7^{+6.7}_{-9.2}$&A\\
Akn 120&H$\beta$&49980--50175&$29.7^{+3.3}_{-5.9}$&B\\
Fairall 9&H$\beta$&50473--50665&$19.4^{+42.1}_{-3.8}$&D\\
Fairall 9&He\,{\sc ii}\,$\lambda1640$&50473--50713&$12.0^{+0.9}_{-3.9}$&C\\
Fairall 9&Ly$\alpha$&50473--50713&$12.1^{+0.5}_{-0.5}$&C\\
Mrk 79&H$\beta$&47838--48044&$25.5^{+2.9}_{-14.4}$&B\\
Mrk 79&H$\beta$&48193--48393&$30.9^{+1.4}_{-2.1}$&A\\
Mrk 79&H$\beta$&48905--49135&$17.2^{+7.3}_{-2.2}$&B\\
Mrk 79&H$\beta$&49996--50220&$43.6^{+1.7}_{-0.8}$&A\\
Mrk 110&H$\beta$&48953--49149&$25.3^{+2.3}_{-13.1}$& B\\
Mrk 110&H$\beta$&49751--49874&$33.9^{+6.1}_{-5.3}$&A\\
Mrk 110&H$\beta$&50010--50262&$21.5^{+2.2}_{-2.1}$&A\\
Mrk 279&H$\beta$&50095--50289&$18.3^{+1.2}_{-1.1}$&A\\
Mrk 290&H$\beta$&54184--54301&$7.7^{+0.7}_{-0.5}$&A\\
Mrk 335&H$\beta$&49156--49338&$15.3^{+3.6}_{-2.2}$&A\\
Mrk 335&H$\beta$&49889--50118&$12.9^{+3.6}_{-5.0}$&A\\
Mrk 509&H$\beta$&47653--50374&$69.9^{+0.3}_{-0.3}$&A\\
Mrk 509&He\,{\sc ii}\,$\lambda4686$&47653--50374&$52.2^{+0.1}_{-0.1}$&D\\
Mrk 590&H$\beta$&48090--48323&$19.0^{+1.8}_{-2.6}$&A\\
Mrk 590&H$\beta$&48848--49048&$19.5^{+2.0}_{-4.0}$&A\\
Mrk 590&H$\beta$&49183--49338&$32.6^{+3.5}_{-8.8}$&B\\
Mrk 590&H$\beta$&49958--50122&$30.9^{+2.5}_{-2.4}$&A\\
Mrk 817&H$\beta$&49000--49212&$20.9^{+2.3}_{-2.3}$&A\\
Mrk 817&H$\beta$&49404--49528&$17.2^{+1.9}_{-2.7}$&A\\
Mrk 817&H$\beta$&49752--49924&$35.9^{+4.8}_{-5.8}$&A\\
Mrk 817&H$\beta$&54200--54330&$10.8^{+1.5}_{-1.0}$&A\\
NGC 3227&H$\beta$&48623--48776&$10.6^{+6.1}_{-6.1}$&A\\
NGC 3227&H$\beta$&54180--54273&$4.4^{+0.3}_{-0.5}$&A\\
NGC 3516&H$\alpha$&47894--48047&$14.0^{+0.7}_{-0.7}$&A\\
NGC 3516&H$\beta$&47894--48047& $6.1^{+0.5}_{-0.7}$&C\\
NGC 3516&H$\beta$&54181--54300& $14.6^{+1.4}_{-1.1}$&A\\
NGC 3783&H$\beta$&48607--48833& $7.3^{+0.3}_{-0.7}$&A\\
NGC 4051&H$\beta$&54180--54311& $2.5^{+0.1}_{-0.1}$&B\\
NGC 4151&H$\beta$&53430--53471& $6.0^{+0.6}_{-0.2}$&A\\
NGC 4593&H$\beta$&53430--53471& $4.5^{+0.7}_{-0.6}$&A\\
NGC 7469&H$\beta$&50237--50295& $11.7^{+0.5}_{-0.7}$&A\\
NGC 7469&Si\,{\sc iv}\,$\lambda1400$&50245--50293&$2.0^{+0.4}_{-0.5}$&A\\
NGC 7469&C\,{\sc iv}\,$\lambda1549$&50245--50293&$10.6^{+0.2}_{-0.2}$&A\\
NGC 7469&He\,{\sc ii}\,$\lambda1640$&50245--50293&$0.8^{+0.2}_{-0.2}$&A\\
PG 0026+129&H$\alpha$&48836--51084&$88.0^{+1.5}_{-3.5}$&B\\
PG 0026+129&H$\beta$&48545--51084&$92.7^{+7.0}_{-0.6}$&C\\
PG 0052+251&H$\alpha$&48837--51084&$157.6^{+2.3}_{-2.8}$&A\\
PG 0052+251&H$\beta$&48461--51084&$149.3^{+4.2}_{-1.8}$&C\\
PG 0052+251&H$\gamma$&48461--51084&$154.9^{+1.9}_{-1.9}$&C\\
PG 0804+761&H$\alpha$&48319--51085&$133.4^{+8.6}_{-4.3}$&C\\
PG 0804+761&H$\beta$&48319--51085&$116.8^{+2.6}_{-7.3}$&A\\
PG 0804+761&H$\gamma$&48319--51085&$71.1^{+46.5}_{-3.1}$&D\\
PG 0844+349&H$\alpha$&48319--51085&$20.8^{+0.4}_{-1.4}$&A\\
PG 0844+349&H$\beta$&48319--51085&$12.2^{+5.2}_{-1.3}$&B\\
PG 0844+349&H$\gamma$&48319--51085&$ 17.7^{+2.5}_{-2.1}$&C\\
PG 0953+414&H$\beta$&48319--50997&$162.2^{+3.5}_{-2.9}$&A\\
PG 0953+414&H$\gamma$&48319--50997&$160.2^{+3.3}_{-56.8}$&D\\
PG 1211+143&H$\alpha$&48319--51000&$76.3^{+0.7}_{-0.5}$&C\\
PG 1211+143&H$\beta$&48319--51000&$73.3^{+0.9}_{-25.4}$&D\\
PG 1211+143&H$\gamma$&48319--51000&$57.7^{+15.8}_{-10.1}$&A\\
PG 1226+023&H$\alpha$&48361--50997&$380.0^{+40.7}_{-6.0}$&B\\
PG 1226+023&H$\beta$&48361--50997&$105.5^{+284.1}_{-5.2}$&D\\
PG 1226+023&H$\gamma$&48361--50997&$263.8^{+9.0}_{-10.6}$&A\\
PG 1229+204&H$\alpha$&48319--50997&$45.7^{+2.8}_{-1.1}$&A\\
PG 1229+204&H$\beta$&48319--50997&$42.8^{+2.3}_{-1.1}$&A
\enddata
\end{deluxetable}
\setcounter{table}{0}
\begin{deluxetable}{lcccc}
\tablecolumns{5}
\tablewidth{0pt}
\tablecaption{--- $Continued$}
\tablehead{
\colhead{ }&
\colhead{ }&
\colhead{Julian Dates}&
\colhead{$\tau_{\rm SPEAR}$}&
\colhead{ }\\
\colhead{Object}&
\colhead{Line}&
\colhead{($-$2400000)}&
\colhead{(days)}&
\colhead{Group}\\
\colhead{(1)}&
\colhead{(2)}&
\colhead{(3)}&
\colhead{(4)}&
\colhead{(5)}
}
\startdata
PG 1307+085&H$\alpha$&49130--51000&$189.1^{+4.6}_{-3.6}$&A\\
PG 1307+085&H$\beta$&48319--51042&$188.8^{+5.7}_{-3.7}$&C\\
PG 1307+085&H$\gamma$&48319--51042&$218.9^{+7.2}_{-124.8}$& D\\
PG 1411+442&H$\alpha$&48319--51038&$59.3^{+10.1}_{-6.7}$&A\\
PG 1411+442&H$\beta$&48319--51038&$53.5^{+13.1}_{-5.3}$&B\\
PG 1426+015&H$\beta$&48334--51042&$161.6^{+6.9}_{-11.1}$&A\\
PG 1617+175&H$\alpha$&48362--51085&$106.9^{+9.8}_{-13.3}$&B\\
PG 1617+175&H$\beta$&48362--51085&$88.2^{+31.0}_{-5.9}$&B\\
PG 2130+099&H$\beta$&54352--54450&$23.2^{+4.4}_{-5.8}$&A\\
PG 2130+099&He\,{\sc ii}\,$\lambda4686$&54352--54450&$32.0^{+3.9}_{-4.5}$&A\\
NGC 5548&H$\beta$&47509--47809&$21.2^{+0.8}_{-1.0}$&A\\
NGC 5548&H$\beta$&47861--48179&$16.3^{+0.8}_{-1.3}$&A\\
NGC 5548&H$\beta$&48225--48534&$15.8^{+2.1}_{-1.1}$&A\\
NGC 5548&H$\beta$&48623--48898&$11.0^{+1.2}_{-1.0}$&A\\
NGC 5548&H$\beta$&48954--49255&$15.3^{+1.4}_{-3.0}$&A\\
NGC 5548&H$\beta$&49309--49636&$10.8^{+1.4}_{-1.0}$&A\\
NGC 5548&H$\beta$&49679--50008&$24.2^{+1.3}_{-0.9}$&A\\
NGC 5548&H$\beta$&50044--50373&$16.1^{+0.3}_{-0.6}$&A\\
NGC 5548&H$\beta$&50434--50729&$16.8^{+0.4}_{-0.2}$&A\\
NGC 5548&H$\beta$&50775--51085&$26.9^{+1.5}_{-2.2}$&A\\
NGC 5548&H$\beta$&51142--51456&$23.8^{+3.1}_{-2.3}$&A\\
NGC 5548&H$\beta$&51517--51791&$8.8^{+1.3}_{-3.9}$&A\\
NGC 5548&H$\beta$&51878--52174&$8.7^{+0.5}_{-0.5}$&B\\
NGC 5548&H$\beta$&54180--54332&$16.3^{+1.0}_{-1.2}$&A
\enddata
\tablecomments{Lag estimates and confidence limits for Groups A and B
are calculated by the single-line fits, while those for Groups C and
D are from the two-line fits.}
\label{tab:lags}
\end{deluxetable}

\begin{figure*}[t]
\epsscale{1.0} \plotone{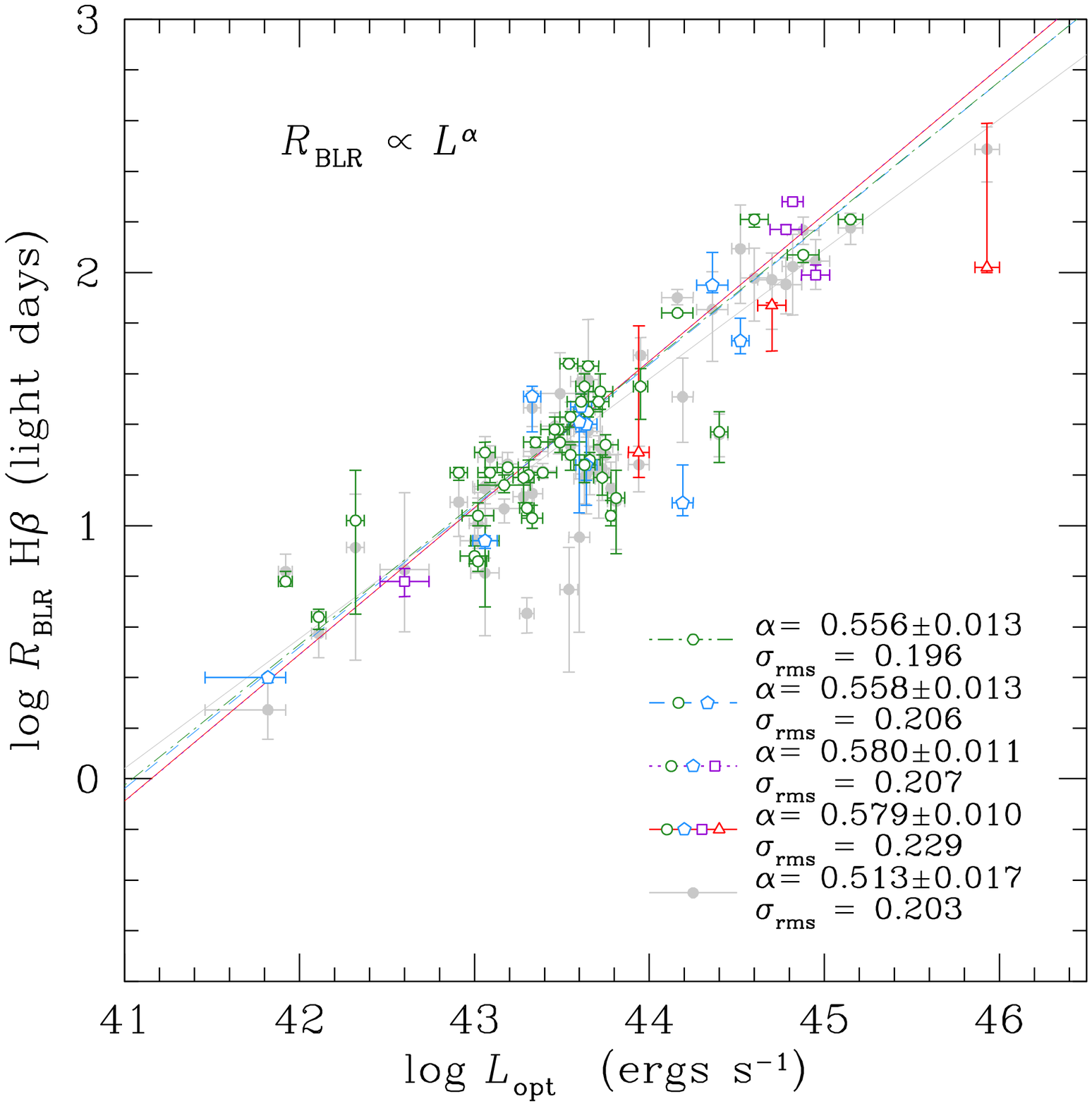}
\caption{The $R_{\rm BLR}$--$L$ relation for H$\beta$. The luminosity
is $\lambda L_{\lambda}(5100\,\text{\rm \AA})$ and the BLR radius is
equivalent to the lag in units of light days. The open symbols and
gray solid circles indicate the measurement from SPEAR method and
from CCF method for the same set of sources, respectively. The gray
solid curve is the fit to the CCF $R_{\rm BLR}$--$L$ relation, while
the rest of the curves are the fits to the SPEAR $R_{\rm BLR}$--$L$
relation, using four subsets of the sources (see
Table~\ref{tab:rl} for details of each fit). The slope of the
fit to the SPEAR $R_{\rm BLR}$--$L$ relation $\alpha$ is steeper than
the CCF relation, but the two are consistent within the
uncertainties. $\sigma_{\rm rms}$ is the rms scatter of each fit.} 
\label{fig:rl} 
\end{figure*}

We performed similar joint analyses for the 21 sources for which we
have multiple emission line light curves and recompile the results
for the H$\beta$ lags, as shown in Figure~\ref{fig:hbetacomp2}.
Fortunately, all the sources whose H$\beta$ lags were found to be
ambiguous in the single-line fits~(i.e., the 7 H$\beta$ lags from
groups \Rmnum{3} in \S\ref{sec:hbeta}) are improved by the two-line
fits, although the degree of improvement varies. We also dropped lag
estimates that were either flagged as unreliable or believed to be
wrong (i.e., the 6 H$\beta$ lags from groups \Rmnum{4} and \Rmnum{5}
in \S\ref{sec:hbeta}) and keep only those objects deemed to give
robust estimates of lag by our method (i.e., the 60 H$\beta$ lags
from groups \Rmnum{1}, \Rmnum{2} and \Rmnum{3}). To illustrate the
quality of the final result for each source, we divide all 60
remaining sources into 4 new groups based on the results of both the
single-line fits in \S~\ref{sec:hbeta} and the two-line fits, using
different symbols for the 4 new groups in
Figure~\ref{fig:hbetacomp2}.
\begin{itemize}
    \item[(A)] The 43 group~\Rmnum{1} light curves from
        \S~\ref{sec:hbeta} with a single unambiguous H$\beta$ lag.
        Seven of the objects  have light curves of lines other than
        H$\beta$ to carry out two-line fits, but they provided little
        gain when the single-line fits already provided good lag
        estimates.
    \item[(B)] The 10 group~\Rmnum{2} sources from \S~\ref{sec:hbeta}
        with a robust H$\beta$ lag estimate but potentially larger
        uncertainties due to the presence of low
        significance~($>3\sigma$) sub-peaks in the lag likelihood
        distribution. Most of those sources do not have the multiple
        line light curves needed to carry out two-line fits.
    \item[(C)] The four group~\Rmnum{3} sources~(NGC~3516, PG~0026,
        PG~0052, and PG~1307) from \S\ref{sec:hbeta} with multiple
        peaks in the single-line lag likelihood distribution where
        the ambiguity is removed by the two-line fits.
    \item[(D)] The three group~\Rmnum{3} sources~(Fairall~9, PG~1211,
        and PG~1226) from \S\ref{sec:hbeta} with multiple peaks in
        the single-line lag likelihood distribution where the
        two-line fits fail to remove the ambiguity. We picked the
        most significant peak as the solution and extended the
        uncertainty to cover all the possible solutions.
\end{itemize}

\tabletypesize{\footnotesize}
\begin{deluxetable*}{lccccccccc}
\tablewidth{0pt}
\tablecaption{BLR Size-Luminosity Relation}
\tablehead{
\colhead{Groups}&
\colhead{ $N$ }&
\colhead{ $C_{\rm SPEAR}$ }&
\colhead{ $\alpha_{\rm SPEAR}$ }&
\colhead{$\chi^2_{\rm SPEAR}$}&
\colhead{$\sigma_{\rm rms}^{\rm SPEAR}$}&
\colhead{ $C_{\rm CCF}$ }&
\colhead{ $\alpha_{\rm CCF}$ }&
\colhead{  $\chi^2_{\rm CCF}$  }&
\colhead{  $\sigma_{\rm rms}^{\rm CCF}$  }\\
\colhead{Included}&
\colhead{ }&
\colhead{(lt-days)}&
\colhead{ }&
\colhead{  }&
\colhead{(dex)}&
\colhead{(lt-days)}&
\colhead{ }&
\colhead{  }&
\colhead{(dex)}\\
\colhead{(1)}&
\colhead{(2)}&
\colhead{(3)}&
\colhead{(4)}&
\colhead{(5)}&
\colhead{(6)}&
\colhead{(7)}&
\colhead{(8)}&
\colhead{(9)}&
\colhead{(10)}
}
\startdata
A,B,C,D&60&$1.36\pm0.01$&$0.579\pm0.010$&8.13&0.229&$1.32\pm0.01$&$0.513\pm0.017$&3.50&0.203\\
A,B,C&57&$1.36\pm0.01$&$0.580\pm0.011$&8.52&0.207&$1.33\pm0.01$&$0.519\pm0.019$&3.54&0.205\\
A,B&53&$1.36\pm0.01$&$0.558\pm0.013$&8.39&0.206&$1.33\pm0.01$&$0.521\pm0.020$&3.81&0.213\\
A&43&$1.36\pm0.01$&$0.556\pm0.013$&9.44&0.196&$1.32\pm0.01$&$0.518\pm0.020$&4.22&0.211
\enddata
\label{tab:rl}
\end{deluxetable*}

Recall that we have dropped the 6 group~\Rmnum{4} and \Rmnum{5}
sources~(IC~4329A, NGC~4593, one season of Mrk~279, one season of
NGC~3227, 3C~120, and PG~1613) out of all 66 H$\beta$ light curves
following the discussion in \S\ref{sec:hbeta}. Green circles, blue
pentagons, dark violet squares, and red triangles correspond to
sources of group A, B, C, and D, respectively. There is general
agreement between the two methods, but also several discrepancies, as
7 of our H$\beta$ lag estimates are inconsistent with the CCF results
given their error estimates.  We marked these sources in
Figure~\ref{fig:hbetacomp2} and now discuss each case individually,

\paragraph{NGC~7469.} We estimate an H$\beta$ lag of
11.7$^{+0.5}_{-0.7}$ days, as opposed to $\tau_{\rm
CCF}=$4.7$^{+0.7}_{-0.8}$. However, if we use a Dirac delta function
for the transfer function instead of a tophat, the estimated time lag
changes to 4.3 days, in agreement with the CCF result. Thus, the
discrepancy originates from the improvement of fit with a tophat
smoothing kernel. The continuum of NGC~7469 was intensively monitored
to search for time lags between the UV and optical
continuum~\citep{collier98}, so its continuum light curve is densely
sampled while the H$\beta$ light curve is much less so. The model has
to smooth the continuum light curve heavily~(i.e., a broad tophat
width) to obtain a good fit, which at the same time shifts the time
lag estimate to a longer value than it would be with a zero
width~(i.e., a delta function). This is suggestive of the continuum
errors being underestimated, or a more realistic transfer
function is required.

\paragraph{Mrk~79~{\rm (years 2 and 4)}.} In both cases, we estimate
larger time lags than the CCF results, although there are sub-peaks
which correspond to the CCF lags. For year 2 (JD47838--48044), while
the CCF centroid gives a lag of 16.4$^{+6.7}_{-6.7}$ days, the CCF
peak estimate is 19$^{+11}_{-12}$ days, more consistent with our
estimate of 30.9$^{+1.4}_{-2.1}$ days.  For year 4 (JD49996--50220),
\cite{peterson04} flagged it as ``unreliable'' for the poor light
curve sampling. Our method shows a dense array of sub-peaks in the
lag likelihood distribution, but the most significant peak is at
43.6$^{+1.7}_{-0.8}$ days.

\paragraph{PG~0844.} As discussed in \S~\ref{sec:corr}, the CCF
estimate of the H$\beta$ lag~(34.4$^{+14.6}_{-14.2}$) for PG~0844 is
likely susceptible to correlated errors, while our method estimates a
lag of 12.2$^{+5.2}_{-1.3}$ days regardless of the value of
correlation coefficient $r$.

\paragraph{PG~0052.} We estimate an H$\beta$ lag of
149.3$^{+4.2}_{-1.8}$ days, as opposed to $\tau_{\rm
CCF}=$103$^{+28.3}_{-27.8}$. The single-line fit shows multiple peaks
and usually one would be inclined to mistrust a peak at the seasonal
alias~(a rest-frame lag of 150 days corresponds to 170 days in the
observed-frame). However, the joint H$\alpha$/H$\beta$ fit clearly
reinforced this solution.

\paragraph{PG~1307.} We estimate an H$\beta$ lag of
188.8$^{+5.7}_{-3.7}$ days, as opposed to $\tau_{\rm CCF}=$
121.9$^{+41.6}_{-53.8}$. The joint H$\alpha$/H$\beta$ fit suppressed
the false peak which corresponds to the $\tau_{\rm CCF}$ lag,
favoring a longer lag that is more consistent with lags of the other
Balmer lines.

\paragraph{PG~1426.} We estimate an H$\beta$ lag of
161.6$^{+6.9}_{-11.1}$ days, as opposed to $\tau_{\rm CCF}=$
103.2$^{+32.5}_{-40.3}$. Similar to PG~0052 and PG~1307, the joint
H$\alpha$/H$\beta$ fit reinforced a solution which is otherwise
susceptible to the seasonal gap effect.

\medskip

We carried out a similar analysis for each data set, including all
emission lines besides H$\beta$, as summarized in
Table~\ref{tab:lags}. Note that in the table we only include 87 light
curves for which we have successfully computed lags.  The object is
identified in column (1). The emission line and its light curve
Heliocentric Julian Date range are listed in columns (2) and (3), respectively.
Column (4) gives the rest-frame time lag estimate from the SPEAR
method, while column (5) indicates the associated
``ambiguity''~(i.e., the group membership) defined above.

\begin{figure*}[!hbt]
\epsscale{1.0} \plotone{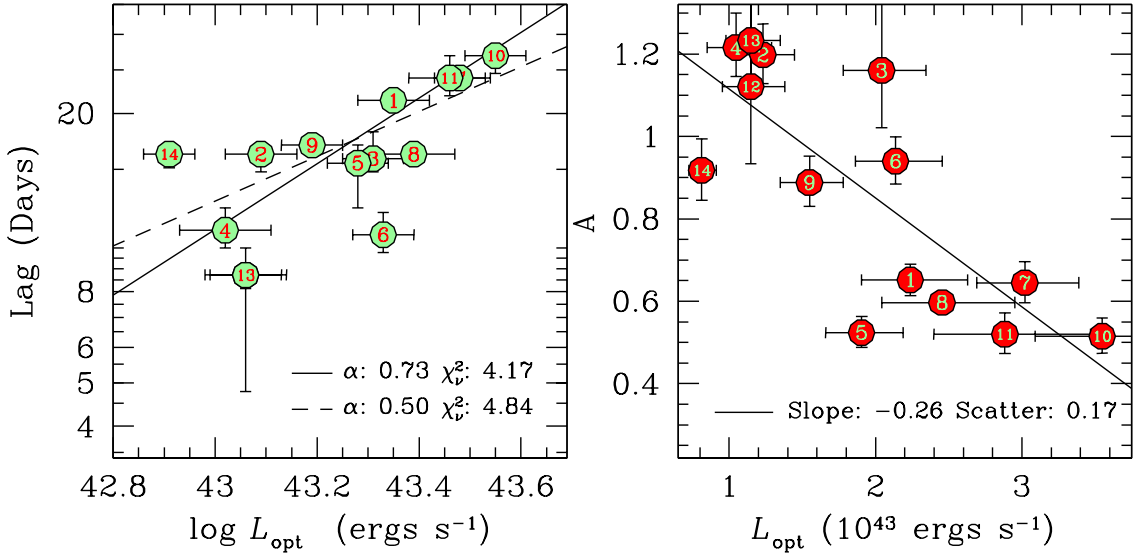}
\caption{Lag~(left) and scaling coefficient~(right) of the H$\beta$
transfer function as a function of continuum luminosity from 14 years
of NGC~5548 data. The best-fit slopes are also reported for each
panel and shown by the black solid lines. The black dashed line in
the left panel is the best fit with a fixed slope of 0.50. The number
inside each solid circle indicates the year of observation for each
light curve starting from Dec. 1988. Note that in the left panel the
point for year 12 is hidden under that of year 13.}
\label{fig:lrs}
\end{figure*}

\section{the $R_{\rm BLR}$--$L$ Relation for H$\beta$}
\label{sec:RLrelation}

With the revised set of H$\beta$ lags, and the starlight-corrected
optical luminosity of each AGN from \cite{bentz09b}, we have
calculated the fit to the $R_{\rm BLR}$--$L$ relationship for our
sample 
\begin{equation}
    R_{\rm BLR} = C  \cdot
    \left(\frac{L}{10^{43.5}\,\rm{ergs\;s}^{-1}}\right)^\alpha
    \label{eqn:rl} 
\end{equation}
and compared it to that based on CCF lags in Figure~\ref{fig:rl}. We
obtained a slope $\alpha = 0.579\pm0.010$ for all the SPEAR lags
regardless of the level of ``ambiguity'' at which we probe the slope.
This slope is slightly steeper than previous estimates, and only
marginally consistent with the na\"{i}ve theoretical prediction of
$\alpha=0.5$. Compared to the CCF-based $R_{\rm BLR}$--$L$ fit of the
same sample of AGNs~(blue filled circles), our $R_{\rm BLR}$--$L$ fit
has a steeper slope but comparable rms scatter $\sigma_{\rm rms}$,
which grows smaller as we use more reliable lags.  Table~\ref{tab:rl}
gives the results from the different fits using the 4 combinations of
groups indicated by column (1). Column (2) gives the number of data
points used in each fit. We fit each combinatorial data set using lag
estimates from both the SPEAR~(columns 3--6) and CCF methods~(columns
7--10). Two parameters in Equation~(\ref{eqn:rl}) are listed in
columns (3) and (4) for SPEAR method, and in column (7) and (8) for
CCF method, respectively. Column (5) and (6) give the $\chi^2/dof$
and rms scatter for our fit, while column (9) and (10) give these
statistics correspondingly for the CCF method. 

Our $R_{\rm BLR}$--$L$ fits have a larger $\chi^2/dof$ than the CCF
ones.  This does not necessarily mean they are poorer fits, because
our lag estimates generally have tighter errorbars than the CCF
estimates. It could indicate that our approach underestimates
uncertainties, that the CCF method overestimates uncertainties, or
that we are not taking into account intrinsic scatter in the $R_{\rm
BLR}$--$L$ relationship. Since most of the group C and D sources are
high-redshift luminous PG objects, the rms scatter for our method
decreases from 0.229 dex to 0.196 dex after dropping them from the
fit. Three outliers from the CCF $R_{\rm BLR}$--$L$
relation~(NGC~7469, years 1 and 4 of Mrk~79) are also the sources
where our lag estimates are inconsistent with CCF results. When we
use our lag estimates, these three CCF outliers lie on the $R_{\rm
BLR}$--$L$ relation, which reduces the rms scatter near $L_{\rm opt}
\sim 10^{43.5}$ ergs s$^{-1}$. Note that there is significant scatter
in the $R_{\rm BLR}$--$L$ relation even for multiple estimates for a
single source, as shown in the left panel of Figure~\ref{fig:lrs} for
NGC~5548.

\section{$R_{\rm BLR}$--$L$ Relation of NGC~5548 Revisited}
\label{sec:breath}
So far, we have carried out our calculations assuming that the
parameters are constant during a season. This is likely true for the
underlying variability process. If we model either the full continuum
light curve or the individual seasons, we find estimates for the
process parameters $\tau_{\rm d}$ and $\hat{\sigma}$ that are
statistically consistent. We do observe lags that vary from season to
season, and these are arguably correlated with luminosity. If so,
they should also be varying within seasons, and we have not accounted
for this. Similarly, we assume the scaling between the continuum and
line fluxes does not vary over a season, although we do observe it to
vary between seasons.

The nearby Seyfert 1 galaxy NGC~5548, with its many continuous years
of monitoring data, serves as an ideal example of an AGN changing its
variability levels from season to season. Figure~\ref{fig:lrs}
illustrates the continuum flux dependence of both the H$\beta$ lag
$\langle \tau \rangle$ and the scaling coefficient $A$ for 14 seasons
of NGC~5548 data. We clearly see trends that the lag increases with
luminosity and the amplitude of the response diminishes. If we fit
the lag, we find a steep slope, $\langle \tau \rangle \propto
L^{0.73\pm0.10}$ that is inconsistent with the expected $\langle \tau
\rangle \propto L^{0.5}$. However, the poor fit ($\chi^2/dof=4.17$)
suggests that either the uncertainties are underestimated or
intrinsic scatter dominates the goodness-of-fit. If we rescale the
uncertainties so that the best-fit model has $\chi^2/dof \equiv 1$,
the flatter $L^{0.5}$ slope is not ruled out, with a $\Delta\chi^2$
of only $0.16$.

\begin{figure}[t] \epsscale{1.0} \plotone{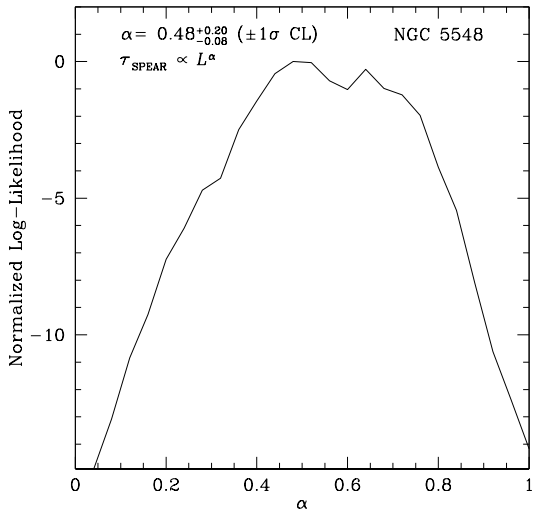}
    \caption{Likelihood distribution of $\alpha$ for 13 years of
    NGC~5548 light curves. The normalized log-likelihood is
    calculated by adding the likelihood distribution functions for
    the 13 individual years together. } \label{fig:breath}
\end{figure}

These problems can be addressed by making the lags and the
line-to-continuum scaling a function of the continuum luminosity. For
the luminosity dependence of lags, the simplest approach would be to
de-lag the line light curve as $\langle \tau \rangle \propto
L^{\alpha}$ instead of shifting the entire light curve by the same
$t_{lag}$, and then optimize the fits over the additional parameter
$\alpha$. Unfortunately, we cannot fit the full NGC~5548 light curve
because the resulting matrix dimensions are impractically high
($K$=3085 data points).  We instead estimate the normalized
likelihood distribution for $\alpha$ in each season and then combined
the likelihoods, as shown in Figure~\ref{fig:breath}~(we did not
include year 13, which was part of the less reliable group B). This
``breathing'' effect is clearly detected, and the logarithmic slope
estimate of $\alpha = 0.44^{+0.20}_{-0.08}$ is consistent with the
na\"{i}ve expectation $\alpha=1/2$ and the $R_{\rm BLR}$--$L$
relation in Figure~\ref{fig:rl}.  Using almost the same set of light
curves from NGC~5548~(we add year 14 and exclude year 13),
\cite{cackett06} find a much shallower slope~($0.1$--$0.46$) with a
luminosity-dependent delay map, in better agreement with the
prediction of photoionization models~($\sim0.23$; \citealt{KG04}).
However, their small correction for the host galaxy starlight may
artificially flatten their estimate of the slope~\citep{bentz07}.
Note that for this experiment we did not make the line-to-continuum
scaling coefficient $A$ a function of continuum luminosity in the
fit. Such a full scale calculation should be carried out using the
complete data set.


\section{Discussion} \label{sec:dis}
We have demonstrated that direct fitting of continuum and line light
curves is a viable approach to measuring reverberation lags,
confirming the initial study of \cite{RK94}. It provides a full
statistical framework for determining time lags and estimating their
uncertainties, including the full contributions from correlated
noise, de-trending and interpolation. In essence, the lags are
determined using a weighted average of all statistically acceptable
models for interpolating the underlying {\it true} light curve. While
we used the assumption that the underlying variable process had an
exponential correlation function corresponding to a damped random
walk, any other statistical process could be substituted. We note,
however, that \cite{kelly09}, \cite{koz10a} and \cite{macleod10} have
found the exponential correlation function to be an excellent model
of quasar light curves, just as we have found here, although we
modeled the light curves in flux rather than magnitude.

Because we are explicitly modeling the light curves, we must include
an explicit model of the transfer function. Here we used a top hat
for simplicity. It includes the simple limits of a delta function and
a uniform thin shell, and is likely a reasonable model for any
single-peaked transfer function given the available data~(see
\citealt{RK94}). As with the model for the variability process, using
an alternative transfer function simply requires computing the
appropriate terms of the covariance matrix. Aside from the case of
NGC~7469 where it seemed to affect the lag estimation, we did not
discuss the tophat width. In general, there is a relatively strong
degeneracy between $\Delta t$, the width, and $A$, the scaling
between the continuum and line light curves. When $\Delta t$ is large
and the continuum is heavily smoothed, the model will try to increase
the variability amplitude by artificially boosting $A$ to re-align
the continuum and line light curves. However, the degeneracy does not
seem to lead to problems in estimating the mean lag unless the line
light curve is very poorly sampled. The traditional CCF method does
not implicitly assume a shape for the transfer function but
calculates the lag as either the barycenter ($\tau_{cent}$) or the
peak ($\tau_{peak}$) of underlying transfer function (convolved with
data).  The difference between the two sometimes can be large and
hard to reconcile unless the transfer function can be modeled
explicitly. For future high-fidelity datasets, our approach should
also have no difficulty constraining the shape of transfer functions.

The most important future path for this method is to simultaneously
fit multiple line components, whether different lines (e.g.,
H$\beta$, H$\alpha$, etc.), velocity sub-components of individual
lines or multiple continuum bands. As long as the overall {\it
ansatz} that all light curves are scaled and smoothed versions of the
continuum holds, combining many light curves with differing lags
means that the lag estimate for any given light curve is now derived
from a better sampled estimate of the continuum variability.  A
second advantage, particularly for attempts to study the velocity
structure of a particular broad line, is that such joint analyses
will correctly infer the covariances between the individual lags.
Current velocity-dependent lags have uncertainties comparable to
their differences~(\citealt{bentz08,bentz10,denney09,denney10}), but
it may be true that these differences actually have a strong
covariances, so that the differences are far more significant than
estimates from analyzing the light curves in isolation. The method
can also allow for luminosity-dependent lags or line-continuum
scaling factors. Also note that while we only use the linear
parameters of the model to remove the light curve means, it is a very
flexible tool for de-trending or cross-calibrating light curves whose
model uncertainties will be fully included in lag estimate.

The most important observational implication of this approach is the
value of measuring multiple lines, especially those with high
ionization potentials.  In our approach, multiple lines with
differing lags allow one to overcome many of the sampling problems
inherent to cross-correlation methodologies. At its simplest, one
light curve can be aliased into a (seasonal) sampling gap, but two
cannot be unless the transfer functions are similar~(i.e., the lines
have similar lags). Given the radial ionization stratification of the
BLR~\citep{peterson93}, the lag difference between two lines is
proportional to the difference in their ionization levels. In this
paper, however, the lines we used for two-line fits are mostly pairs
of two Balmer lines, which have similarly low ionization levels.
Thus, the observational goal should be to obtain data for multiple
lines with a broad range of ionization potentials. Indeed, with a
wide variety of emission-line lines, it is in principle possible to
combine the reverberation results with photoionization equilibrium
modeling to highly constrain the geometry and physics of the BLR
\citep{HKG03}.

The only significant algorithmic challenge comes from the
$\mathcal{O}(K^3)$ scaling of the computational cost with the number
of data points $K$.  Unfortunately, the reverberation mapping problem
is very different from simply using the damped random walk to model
the continuum light curves, where we can take advantage of the
particular structure of the covariance matrix to calculate the
necessary matrix inversions in $\mathcal{O}(K)$ operations.  Since
the expensive matrix inversion is required for each likelihood
calculation, it becomes difficult to analyze large data sets,
particularly if the number of parameters also increases greatly as in
a full simultaneous model of lags as a function of line velocity.
These problems can be addressed using hyper--threaded or parallel
versions of the underlying algorithm.

\section*{Acknowledgements} We thank Kelly D. Denney and Catherine J.
Grier for kindly providing some of the light curves. Thanks also to
Misty C. Bentz for her starlight corrected AGN luminosities. CSK is
supported by NSF Grant AST-0708082 and AST-1009756. BMP is supported
by NSF Grant AST-1008882 and YZ is supported by an OSU Distinguished
University Fellowship.


\appendix
\section{Covariance Matrix of the Correlation Functions}

The expressions for the covariance matrices used in this paper and
the accompanying code assume that the transfer function is a simple
top hat,
\begin{equation}
    \Psi(t-t') = A \left(t_2-t_1\right)^{-1} \quad\hbox{for}\quad t_1 \leq t-t'
    \leq t_2.
\label{app:tf} 
\end{equation}
For this transfer function, we can analytically calculate the
correlation functions in Equation~(\ref{eqn:lc}), (\ref{eqn:lauto})
and (\ref{eqn:lcross}), respectively.

\subsection{The Covariance Matrix Between the Continuum and One Line}
The covariance between continuum $s_c(t)$ at $t_j$ and line $s_l(t)$
at $t_i$ with transfer function defined as in Equation~(\ref{app:tf})
is
\begin{equation}
    \langle s_c(t_j) s_l(t_i)\rangle = \tau_{\rm d} \sigma^2 A
    \left\{ 
    \begin{array}{l l}
        e^{-t_L/\tau_{\rm d}} - e^{-t_H/\tau_{\rm d}} & \quad \mbox{if $t_L>0$}\\
        e^{{t_H}/{\tau_{\rm d}}} - e^{{t_L}/{\tau_{\rm d}}} & \quad \mbox{if $t_H<0$}\\
        2 - e^{{t_L}/{\tau_{\rm d}}} - e^{-{t_H}/{\tau_{\rm d}}} & \quad \mbox{if
        $t_L\leq 0 \leq t_H$},\\
     \end{array} 
     \right.
    \label{app:cl}
\end{equation}
where $t_L \equiv t_i - t_j - t_2$ and $t_H \equiv t_i - t_j - t_1$.

\subsection{The Covariance Matrix Between Two Lines}

Consider the case when the first line $s_l(t)$ has transfer function
$\Psi(t-t')$ as defined in Equation~(\ref{app:tf}) and the other line
$s'_l(t)$ has transfer function $\Psi'(t-t')$ 
\begin{equation}
    \Psi'(t-t') = B \left(t_4-t_3\right)^{-1} \quad\hbox{for}\quad t_3 \leq t-t'
    \leq t_4,
\end{equation}
where $t_4-t_3 \leq t_2-t_1$. The covariance between line $s_l(t)$ at
time $t_i$ and  line $s'_l(t)$ at time $t_j$
(Equation~\ref{eqn:lcross}) is
\begin{equation}
    \langle s_l(t_i) s'_l(t_j)\rangle = \tau_{\rm d}^2 \sigma^2 A\,B
    \left\{
    \begin{array}{l l}
        e^{{-|t_L|}/{\tau_{\rm d}}} + e^{{-|t_H|}/{\tau_{\rm d}}} -
        e^{{-|t_{M1}|}/{\tau_{\rm d}}} - e^{{-|t_{M2}|}/{\tau_{\rm d}}} + 
        \left\{
        \begin{array}{l l}
            {2\,t_H}/{\tau_{\rm d}} &\quad \mbox{if $t_{M2} \leq 0 < t_H$} \\
            {2\,(t_4-t_3)}/{\tau_{\rm d}} &\quad \mbox{if $t_{M2} \leq 0 < t_H$} \\
            {-2\,t_L}/{\tau_{\rm d}} &\quad \mbox{if $t_{L} \leq 0 < t_{M1}$} \\
        \end{array}\right.&\\
        e^{{-|t_L|}/{\tau_{\rm d}}} + e^{{-|t_H|}/{\tau_{\rm d}}} -
        e^{{-|t_{M1}|}/{\tau_{\rm d}}} - e^{{-|t_{M2}|}/{\tau_{\rm d}}}\qquad \mbox{if
        $t_L>0$ or $t_H<0$}, & \\
    \end{array} 
    \right.
    \label{app:ll}
\end{equation}
where 
\begin{eqnarray}
t_L &\equiv& (t_i-t_j)-(t_2-t_3), \nonumber \\ 
t_{M1} &\equiv& (t_i-t_j)-(t_2-t_3),\nonumber \\
t_{M2} &\equiv& (t_i-t_j)-(t_1-t_3), \nonumber\\
\quad \mbox{and} \quad t_{H} &\equiv& (t_i-t_j)-(t_1-t_4).
\end{eqnarray}

By definition, the covariance for the autocorrelation of line
$s_l(t)$ between time $t_i$ and $t_j$ (Equation~\ref{eqn:lauto}) can
be obtained by equating $\Psi'(t-t')$ with $\Psi(t-t')$ so that $B
\equiv A$, $t_3\equiv t_1$ and $t_4\equiv t_2$.



\begin{thebibliography}{20}

\bibitem[Alexander(1997)]{alexander97} Alexander, T.\ 1997, 
in Astronomical Time Series, ed.\ Maoz, D., Sternberg, A.,
\& Leibowitz, E.~M. (Dordrecht: Kluwer), p.\ 163 

\bibitem[Bentz et al.(2006)]{bentz06a} Bentz, M.~C., Peterson, 
B.~M., Pogge, R.~W., Vestergaard, M., \& Onken, C.~A.\ 2006, \apj, 644, 133 


\bibitem[{Bentz {et~al.}(2006)Bentz, Denney, Cackett, Dietrich, Fogel, Ghosh,
    Horne, Kuehn, Minezaki, Onken, Peterson, Pogge, Pronik, Richstone,
    Sergeev, Vestergaard, Walker, \& Yoshii}]{bentz06b} Bentz, M.~C., {et~al.}
    2006, \apj, 651, 775

\bibitem[Bentz et al.(2007)]{bentz07} Bentz, M.~C., et al.\ 2007, \apj, 662,
    205 

\bibitem[Bentz et al.(2008)]{bentz08} Bentz, M.~C., et al.\ 
    2008, \apjl, 689, L21 

\bibitem[Bentz et al.(2009)]{bentz09a} Bentz, M.~C., Peterson, B.~M., Pogge,
    R.~W., \& Vestergaard, M.\ 2009, \apjl, 694, L166 

\bibitem[Bentz et al.(2009)]{bentz09b} Bentz, M.~C., Peterson, 
    B.~M., Netzer, H., Pogge, R.~W., \& Vestergaard, M.\ 2009, \apj, 697, 160 

\bibitem[Bentz et al.(2010)]{bentz10} Bentz, M.~C., et al.\ 
    2010, ApJ, 720, 46

\bibitem[Blandford \& McKee(1982)]{BM82} Blandford, R.~D., \& McKee, C.~F.\
    1982, \apj, 255, 419 

\bibitem[Cackett \& Horne(2006)]{cackett06} Cackett, E.~M., \& Horne, K.\
    2006, \mnras, 365, 1180 

\bibitem[Clavel et al.(1991)]{clavel91} Clavel, J., et al.\ 
1991, \apj, 366, 64 

\bibitem[{Collier \& Peterson(2001)}]{CB01} Collier, S., \& Peterson, B.~M.
    2001, \apj, 555, 775

\bibitem[Collier et al.(1998)]{collier98} Collier, S.~J., et al.\ 1998, \apj,
    500, 162 

\bibitem[Denney et al.(2006)]{denney06} Denney, K.~D., et al.\ 2006,
\apj, 653, 152 


\bibitem[Denney et al.(2009)]{denney09} Denney, K.~D., et al.\ 
    2009, \apjl, 704, L80 

\bibitem[Denney et al.(2010)]{denney10} Denney, K.~D., et al.\ 2010,
    ApJ, 721, 715 

\bibitem[Edelson \& Krolik(1988)]{EK88} Edelson, R.~A., \& Krolik, J.~H.\
    1988, \apj, 333, 646 

\bibitem[Ferrarese \& Merritt(2000)]{FM00} Ferrarese, L., \& Merritt, D.\
    2000, \apjl, 539, L9 

\bibitem[Ferrarese et al.(2001)]{ferrarese01} Ferrarese, L., Pogge, R.~W.,
    Peterson, B.~M., Merritt, D., Wandel, A., \& Joseph, C.~L.\ 2001, \apjl,
    555, L79 

\bibitem[Gaskell \& Sparke(1986)]{GS86} Gaskell, C.~M., \& Sparke, L.~S.\
    1986, \apj, 305, 175 

\bibitem[Gaskell \& Peterson(1987)]{GP87} Gaskell, C.~M., \& Peterson, B.~M.\
    1987, \apjs, 65, 1 

\bibitem[Gebhardt et al.(2000)]{gebhardt00a} Gebhardt, K., et al.\ 2000,
    \apjl, 539, L13

\bibitem[Gebhardt et al.(2000)]{gebhardt00b} Gebhardt, K., et al.\ 2000,
    \apjl, 543, L5 

\bibitem[Graham et al.(2011)]{graham11} Graham, A.W., Onken, C.A.,
  Athanassoula, E. \& Combes, F. 2011, submitted to MNRAS
  (arXiv:1007.3834)

\bibitem[{Grier {et~al.}(2008)Grier, Peterson, Bentz, Denney, Eastman,
    Dietrich, Pogge, Prieto, {DePoy}, Assef, Atlee, Bird, Eyler, Peeples,
    Siverd, Watson, \& Yee}]{grier08} Grier, C.~J., {et~al.} 2008, \apj, 688,
    837

\bibitem[G{\"u}ltekin et al.(2009)]{gultekin09} G{\"u}ltekin, K., et al.\
    2009, \apj, 698, 198 

\bibitem[{Hastings(1970)}]{hastings70} Hastings, W.~K. 1970, Biometrika, 57,
    97

\bibitem[Hopkins \& Hernquist(2006)]{HH06} Hopkins, P.~F., \& Hernquist, L.\
    2006, \apjs, 166, 1 

\bibitem[Horne, Korista, \& Goad(2003)]{HKG03} Horne, K., Korista, K.~T.,
\& Goad, M.~R. 2003, MNRAS, 339, 367

\bibitem[Horne et al.(2004)]{horne04} Horne, K., Peterson, B.~M., Collier,
    S.~J., \& Netzer, H.\ 2004, \pasp, 116, 465 

\bibitem[Kaspi et al.(1996)]{kaspi96} Kaspi, S., Smith, P.~S., 
Maoz, D., Netzer, H., \& Jannuzi, B.~T.\ 1996, \apjl, 471, L75 

\bibitem[{Kaspi {et~al.}(2000)Kaspi, Smith, Netzer, Maoz, Jannuzi, \&
    Giveon}]{kaspi00} Kaspi, S., Smith, P.~S., Netzer, H., Maoz, D., Jannuzi,
    B.~T., \& Giveon, U.  2000, \apj, 533, 631

\bibitem[{Kaspi {et~al.}(2005)Kaspi, Maoz, Netzer, Peterson, Vestergaard, 
\& Jannuzi}]{kaspi05} Kaspi, S., Maoz, D, Netzer, H., Peterson, B.M.,
Vestergaard, M., Jannuzi, B.T. 2005, \apj, 629, 61

\bibitem[{Kelly {et~al.}(2009)Kelly, Bechtold, \& Siemiginowska}]{kelly09}
    Kelly, B.~C., Bechtold, J., \& Siemiginowska, A. 2009, \apj, 698, 895

\bibitem[Kelly et al.(2010)]{kelly10} Kelly, B.~C., Vestergaard, M., Fan, X.,
    Hopkins, P., Hernquist, L., \& Siemiginowska, A.\ 2010, ApJ, 719, 1315

\bibitem[Kollmeier et al.(2006)]{kollmeier06} Kollmeier, J.~A., et al.\ 2006,
    \apj, 648, 128 

\bibitem[{{Koz{\l}owski} \& {Kochanek}(2009)}]{KK09} {Koz{\l}owski}, S., \&
    {Kochanek}, C.~S. 2009, \apj, 701, 508

\bibitem[Koz{\l}owski et al.(2010)]{koz10a} Koz{\l}owski, S., et al.\ 2010,
    \apj, 708, 927 

\bibitem[Korista \& Goad(2004)]{KG04} Korista, K.~T., \& Goad, M.~R.\ 2004,
    \apj, 606, 749 

\bibitem[Kormendy \& Richstone(1995)]{KR95} Kormendy, J., \& Richstone, D.\
    1995, \araa, 33, 581 

\bibitem[Krolik(1999)]{krolik99} Krolik, J.~H.\ 1999, Active Galactic Nuclei:
    From the Central Black Hole to the Galactic
    Environment~(Princeton~,~N.~J.: Princeton University Press)  

\bibitem[Krolik et al.(1991)]{krolik91} Krolik, J.~H., Horne, 
K., Kallman, T.~R., Malkan, M.~A., Edelson, R.~A., 
\& Kriss, G.~A.\ 1991, \apj, 371, 541 

\bibitem[Labita et al.(2006)]{labita06}Labita, M., Treves, A.,
  Falomo, R., \& Uslenghi, M. 2006, MNRAS, 373, 551

\bibitem[MacLeod et al.(2010)]{macleod10} MacLeod, C.~L., et al.\ 2010,
ApJ, 721, 1014

\bibitem[Magorrian et al.(1998)]{magorrian98} Magorrian, J., et al.\ 1998,
    \aj, 115, 2285 

\bibitem[Marconi et al.(2008)]{marconi08} Marconi, A., Axon, 
D.~J., Maiolino, R., Nagao, T., Pastorini, G., Pietrini, P., Robinson, A., 
\& Torricelli, G.\ 2008, \apj, 678, 693 

\bibitem[Marconi et al.(2009)]{marconi09} Marconi, A., Axon, 
D.~J., Maiolino, R., Nagao, T., Pietrini, P., Risaliti, G., Robinson, A., 
\& Torricelli, G.\ 2009, \apjl, 698, L103 

\bibitem[McLure \& Jarvis(2002)]{mclure02} 
McLure, R.~J., \& Jarvis, M.~J.\ 2002, \mnras, 337, 109 

\bibitem[McLure \& Jarvis(2004)]{mclure04} 
McLure, R.~J., \& Jarvis, M.~J.\ 2004, \mnras, 353, L45 

\bibitem[{{Metropolis} {et~al.}(1953){Metropolis}, {Rosenbluth},
    {Rosenbluth}, {Teller}, \& {Teller}}]{MRRTT53} {Metropolis}, N.,
    {Rosenbluth}, A.~W., {Rosenbluth}, M.~N., {Teller}, A.~H., \& {Teller},
    E. 1953, \jcp, 21, 1087

\bibitem[Nelson et al.(2004)]{nelson04} Nelson, C.~H., Green, R.~F., Bower,
    G., Gebhardt, K., \& Weistrop, D.\ 2004, \apj, 615, 652 

\bibitem[Netzer \& Marziani(2010)]{NM10} 
Netzer, H., \& Marziani, P.\ 2010, \apj, 724, 318 

\bibitem[Netzer(2009)]{netzer09} Netzer, H.\ 2009, \apj, 695, 
793 

\bibitem[Osterbrock(1989)]{osterbrock89} Osterbrock, D.~E.\ 1989,
    Astrophysics of Gaseous Nebulae and Active Galactic Nuclei~(Mill
    Valley~CA: University Science Books)

\bibitem[Onken et al.(2004)]{onken04} Onken, C.~A., Ferrarese, L., Merritt,
    D., Peterson, B.~M., Pogge, R.~W., Vestergaard, M., \& Wandel, A.\ 2004,
    \apj, 615, 645 


\bibitem[Peng et al.(2006)]{peng06} Peng, C.~Y., Impey, C.~D., Ho, L.~C.,
    Barton, E.~J., \& Rix, H.-W.\ 2006, \apj, 640, 114 

\bibitem[Peterson(1993)]{peterson93} Peterson, B.~M.\ 1993, \pasp, 105, 247 

\bibitem[Peterson(1997)]{peterson97} Peterson, B.~M.\ 1997, An Introduction
    to Active Galactic Nuclei, (Cambridge: Cambridge
    University Press)

\bibitem[Peterson \& Wandel(1999)]{PW99} 
Peterson, B.~M., \& Wandel, A.\ 1999, \apjl, 521, L95 

\bibitem[Peterson \& Wandel(2000)]{PW00} Peterson, B.~M., \& Wandel, A.\
    2000, \apjl, 540, L13 

\bibitem[Peterson(2001)]{peterson01} Peterson, B.~M.\ 2001, Advanced Lectures
    on the Starburst-AGN Connection, ed. I.~Aretxaga, D.~Kunth, \&
    R.~M{\'u}jica (Singapore: World Scientific), p.3

\bibitem[{{Peterson}(2008)}]{peterson08} {Peterson}, B.~M. 2008, New
    Astronomy Review, 52, 240

\bibitem[Peterson et al.(1991)]{peterson91} Peterson, B.~M., et 
al.\ 1991, \apj, 368, 119 


\bibitem[{Peterson {et~al.}(1998)Peterson, Wanders, Bertram, Hunley, Pogge,
    \& Wagner}]{peterson98} Peterson, B.~M., Wanders, I., Bertram, R.,
    Hunley, J.~F., Pogge, R.~W., \& Wagner, R.~M. 1998, \apj, 501, 82

\bibitem[{Peterson {et~al.}(2004)Peterson, Ferrarese, Gilbert, Kaspi, Malkan,
    Maoz, Merritt, Netzer, Onken, Pogge, Vestergaard, \& Wandel}]{peterson04}
    Peterson, B.~M., {et~al.} 2004, \apj, 613, 682

\bibitem[{Press {et~al.}(1992)Press, Rybicki, \& Hewitt}]{PRH92} Press,
    W.~H., Rybicki, G.~B., \& Hewitt, J.~N. 1992, \apj,
    385, 404

\bibitem[{{Press} {et~al.}(1992){Press}, {Teukolsky}, {Vetterling}, \&
    {Flannery}}]{nr} {Press}, W.~H., {Teukolsky}, S.~A., {Vetterling}, W.~T.,
    \& {Flannery}, B.~P.  1992, {Numerical Recipes in FORTRAN. The Art of
    Scientific Computing}, (Cambridge: Cambridge University Press)

\bibitem[Rauch \& Blandford(1991)]{RB91} Rauch, K.~P., \& Blandford, R.~D.\
    1991, \apjl, 381, L39 

\bibitem[{{Rybicki} \& {Kleyna}(1994)}]{RK94} {Rybicki}, G.~B., \& {Kleyna},
    J.~T. 1994, in ASP Conf. Ser. 69, Reverberation Mapping of the Broad-Line
    Region in Active Galactic Nuclei, ed. P.~M.~Gondhalekar, K.~Horne, \&
    B.~M.~Peterson (San Francisco: ASP), p.~85

\bibitem[{Rybicki \& Press(1992)}]{RP92} Rybicki, G.~B., \& Press, W.~H.
    1992, \apj, 398, 169

\bibitem[{Rybicki \& Press(1995)}]{RP95} ---. 1995, Physical Review Letters,
    74, 1060

\bibitem[Steinhardt \& Elvis(2010)]{SE10} Steinhardt, C.~L., \& Elvis, M.\
    2010, \mnras, 402, 2637 

\bibitem[Shankar et al.(2009)]{shankar09} Shankar, F., Weinberg, D.~H., \&
    Miralda-Escud{\'e}, J.\ 2009, \apj, 690, 20 

\bibitem[Shen et al.(2008)]{shen08} Shen, Y., Greene, J.~E., Strauss, M.~A.,
Richards, G.~T., \& Schneider, D.P. 2008, \apj, 680, 169
 
\bibitem[Tremaine et al.(2002)]{tremaine02} Tremaine, S., et al.\ 2002, \apj,
    574, 740 

\bibitem[{{Udalski} {et~al.}(2008){Udalski}, {Szymanski}, {Soszynski}, \&
    {Poleski}}]{udalski08} {Udalski}, A., {Szymanski}, M.~K., {Soszynski},
    I., \& {Poleski}, R. 2008, Acta Astronomica, 58, 69

\bibitem[Vestergaard(2002)]{vestergaard02} Vestergaard, M.\ 2002, 
\apj, 571, 733 

\bibitem[Vestergaard \& Peterson(2006)]{vestergaard06} 
Vestergaard, M., \& Peterson, B.~M.\ 2006, \apj, 641, 689 

\bibitem[{Wandel {et~al.}(1999)Wandel, Peterson, \& Malkan}]{wandel99}
    Wandel, A., Peterson, B.~M., \& Malkan, M.~A. 1999, \apj, 526, 579

\bibitem[Wanders et al.(1993)]{wanders93} Wanders, I., et al.\ 1993, \aap,
    269, 39


\bibitem[Wambsganss(2006)]{wambsganss06} Wambsganss, J.\ 2006, Gravitational
    Lensing: Strong, Weak and Micro, Saas-Fee Advanced Courses, Volume
    33. (Berlin: Springer-Verlag) p.~453

\bibitem[Welsh(1999)]{welsh99} Welsh, W.~F.\ 1999, \pasp, 111, 
1347 


\bibitem[White \& Peterson(1994)]{WP94} White, R.~J., \& Peterson, B.~M.\
    1994, \pasp, 106, 879 

\bibitem[Woo et al.(2010)]{woo10} Woo, J.-H., et al. 2010, ApJ,
    716, 269

\end{thebibliography}

%
\end{document}